%% file: main.tex
\PassOptionsToPackage{x11names,dvipsnames,table}{xcolor}
\documentclass[twocolumn,5p]{elsarticle}
\usepackage[bitstream-charter]{mathdesign}

\usepackage{hyperref}
\input{preamble}

\let\today\relax
\makeatletter
\def\ps@pprintTitle{%
    \let\@oddhead\@empty
    \let\@evenhead\@empty
    \def\@oddfoot{\footnotesize\itshape
         {} \hfill\today}%
    \let\@evenfoot\@oddfoot
    }
\makeatother

\begin{document}

\begin{frontmatter}

    \title{Function-specific Scheduling Policies in Cloud-Edge, Multi-Region Serverless Systems}

    \author[unibo]{Giuseppe De Palma\corref{cor1}}
    \ead{giuseppe.depalma2@unibo.it}
    \cortext[cor1]{Corresponding author}
    \author[unibo,inria]{Saverio Giallorenzo}
    \ead{saverio.giallorenzo2@unibo.it}
    \author[sdu]{Jacopo Mauro}
    \ead{mauro@imada.sdu.dk}
    \author[unibo]{Matteo Trentin}
    \ead{matteo.trentin2@unibo.it}
    \author[unibo,inria]{Gianluigi Zavattaro}
    \ead{gianluigi.zavattaro@unibo.it}

    \address[unibo]{Universit\`a di Bologna, Dipartimento di Informatica --- Scienza e Ingegneria Mura Anteo Zamboni 7, 40126, Bologna, Italy}
    \address[inria]{INRIA, Sophia-Antipolis, France}
    \address[sdu]{University of Southern Denmark, Department of Mathematics and Computer Science (IMADA), University of Southern Denmark, Campusvej 55, Odense M, 5230, Denmark}

    \begin{abstract}
        %Serverless applications deployed over the could-edge continuum or over multiple regions
        Cloud-edge serverless applications or serverless deployments spanning multiple regions %deployments 
        introduce the
        need to govern the scheduling of functions to satisfy  their functional
        constraints or avoid performance degradation. For instance, functions may require to be allocated to
        specific private (edge) nodes that
        have access to specialised resources or to nodes with low latency to access a certain database to decrease the overall latency of the application.

        State-of-the-art serverless platforms do not support directly the implementation of
        topological constraints on the scheduling of functions.
        % A solution is to
        % implement topological constraints with scenario-specific platform deployments.
        % However, this practice is fragile (e.g., difficult to adapt to changing business
        % needs) and (even worse) it makes it difficult to check that the resulting
        % scheduling would respect the constraints since its implementation is
        % intermixed with the other deployment details.
        %
        % Considering these constraints when scheduling functions leads to
        % sensible performance improvements, e.g., minimising failures, loading
        % times or data-access latencies. This issue becomes more pressing when
        % considered in the emerging multi-cloud and edge-cloud-continuum systems,
        % where only specific nodes can access specialised, local resources. 
        %
        We address this problem %from a linguistic perspective. We
        by presenting a declarative
        language for defining topology-aware, function-specific serverless scheduling policies, called \tapp. Given a \tapp script, a compatible
        serverless scheduler can enforce different, co-existing topological constraints
        without requiring ad-hoc platform deployments. We prove our approach feasible by
        implementing a \tapp-based serverless platform as an extension of the Apache
        OpenWhisk serverless platform. We show that, compared to vanilla OpenWhisk, our extension does not negatively impact the performance of generic, non-topology-bound serverless scenarios, while it increases the performance of topology-bound ones.

    \end{abstract}


\end{frontmatter}

\input{intro}

\input{preliminaries}

\input{implementation}

\input{benchmark}
\input{related}

\input{conclusion}

\bibliography{biblio}

\end{document}

%% file: preamble.tex
\usepackage[english]{babel}
\usepackage[utf8]{inputenc}
\usepackage{graphicx}
\usepackage{newlfont}
\usepackage{amssymb}
\usepackage[bbgreekl]{mathbbol}
\usepackage{latexsym}
\usepackage{amsthm}
\usepackage{balance}
\usepackage{booktabs}
\usepackage{inconsolata}
\usepackage{listings}
\usepackage{adjustbox}

\usepackage{pgfplotstable}
\pgfplotsset{compat=1.16}
\usepackage{caption}
\usepackage[caption=false]{subfig}
\usepackage{listofitems}
\usepackage{xspace}

\usepackage{tikz}
\usetikzlibrary{arrows}
\usepackage{amsmath}
\usepackage[capitalise,noabbrev]{cleveref}
\usepackage{multirow}
\usetikzlibrary{decorations.pathreplacing,positioning,arrows.meta,backgrounds}
\usepackage[
  disable,     % hide all todos
  textwidth=\marginparwidth,
  colorinlistoftodos,
  prependcaption,
  linecolor=green,backgroundcolor=green!25,bordercolor=green
]{todonotes}
\makeatletter
\if@todonotes@disabled
\else % todonotes is enabled, increase paper size to get bigger margins
  \newlength{\increase}
  \paperwidth=\dimexpr \paperwidth + \increase\relax
  \oddsidemargin=\dimexpr\oddsidemargin + .5\increase\relax
  \evensidemargin=\dimexpr\evensidemargin + .6\increase\relax
  \marginparwidth=\dimexpr \marginparwidth + .75\increase\relax
  \let\tmptitle\title
  \renewcommand{\title}[1]{\tmptitle{#1{\tiny\normalfont \makebox[10em]{\colorbox{red!20}{todos enabled}}}}}
\fi
\makeatother

\newcommand{\Coloneqq}{::=}
\newcommand{\newFrag}[1]{\text{#1}}

\pgfplotstableset{
  col sep=comma,
  every head row/.style={
      before row=\hline,
      after row=\hline\hline},
  string type,
  column type={|c},
  every last column/.style={column type={|c|}},
}
\pgfplotstableset{
  highlightrow/.style={
      postproc cell content/.append code={
          \count0=\pgfplotstablerow
          \advance\count0 by1
          \ifnum\count0=#1
            \pgfkeysalso{@cell content/.add={\bf}}
          \fi
        },
    },
}

\pgfplotstableset{NormalStyle/.append style={
      after row={\hline},
      postproc cell content/.append code={
          \ifodd\pgfplotstablerow\relax
          \else
            \pgfkeysalso{@cell content={\cellcolor[gray]{0.9}##1}}%
          \fi
        }
    }}
\pgfplotstableset{ResultsStyle/.append style={
columns={-,Vanilla,Modified - default,isolated,min_memory,shared},
columns/min_memory/.style={
    column name=min\_memory
  },
every column/.style={
column type={|>{\centering\arraybackslash}p{.15\linewidth}},
string type
},
columns/-/.style={
column type={|>{\centering\arraybackslash}p{.09\linewidth}|},
string type
},
}}
\pgfplotstableset{ResultsStyleData/.append style={
columns={-,Vanilla,Modified - default,isolated,min_memory,shared,shared - tag},
columns/min_memory/.style={
    column name=min\_memory
  },
every column/.style={
column type={|>{\centering\arraybackslash}p{.12\linewidth}},
string type
},
columns/-/.style={
column type={|>{\centering\arraybackslash}p{.09\linewidth}|},
string type
},
}}

\newcommand\YAMLcolonstyle{\color{BrickRed}\ttfamily}
\newcommand\YAMLkeystyle{\color{black}\ttfamily}
\newcommand\YAMLvaluestyle{\color{blue}\ttfamily}

\makeatletter

\newcommand\language@yaml{yaml}

\expandafter\expandafter\expandafter\lstdefinelanguage
\expandafter{\language@yaml}
{
keywords={workers,controller,strategy,followup,invalidate,topology_tolerance},
keywordstyle=\ttfamily\color{RoyalBlue},
basicstyle=\YAMLkeystyle,                                 % assuming a key comes first
tabsize=1,
numberstyle=\footnotesize,
sensitive=false,
comment=[l]{\#},
morecomment=[s]{/*}{*/},
commentstyle=\color{gray}\ttfamily,
stringstyle=\YAMLvaluestyle\ttfamily,
moredelim=[l][\color{orange}]{\&},
moredelim=**[is][\YAMLvaluestyle]{~}{~},
morestring=[b]',
morestring=[b]",
literate =    {---}{{\ProcessThreeDashes}}3
{>}{{\textcolor{red}\textgreater}}1
{|}{{\textcolor{red}\textbar}}1
{\ -\ }{{\mdseries\ -\ }}3
{:}{{{\YAMLcolonstyle{:}}}}1
}

\lst@AddToHook{EveryLine}{\ifx\lst@language\language@yaml\YAMLkeystyle\fi}
\makeatother

\newcommand\JSONnumbervaluestyle{\color{blue}}
\newcommand\JSONstringvaluestyle{\color{red}}

\newif\ifcolonfoundonthisline

\makeatletter

\lstdefinestyle{json}
{
  showstringspaces    = false,
  keywords            = {false,true},
  alsoletter          = 0123456789.,
  morestring          = [s]{"}{"},
  stringstyle         = \ifcolonfoundonthisline\JSONstringvaluestyle\fi,
  MoreSelectCharTable =%
  \lst@DefSaveDef{`:}\colon@json{\processColon@json},
  basicstyle          = \ttfamily,
  keywordstyle        = \ttfamily\bfseries,
}

\newcommand\processColon@json{%
  \colon@json%
  \ifnum\lst@mode=\lst@Pmode%
    \global\colonfoundonthislinetrue%
  \fi
}

\lst@AddToHook{Output}{%
  \ifcolonfoundonthisline%
    \ifnum\lst@mode=\lst@Pmode%
      \def\lst@thestyle{\JSONnumbervaluestyle}%
    \fi
  \fi
  \lsthk@DetectKeywords% 
}

\lst@AddToHook{EOL}%
{\global\colonfoundonthislinefalse}

\makeatother

\newcommand\ProcessThreeDashes{\llap{\color{cyan}\mdseries-{-}-}}

\definecolor{E}{RGB}{51,153,255}

\definecolor{S}{RGB}{6,153,0}

\definecolor{B}{RGB}{153,0,0}

\newcommand{\hl}[1]{{\color{RoyalBlue}\texttt{#1}}}
\newcommand{\hlopt}[1]{{\color{blue}\texttt{#1}}}
\newcommand{\hlstr}[1]{{\color{BrickRed}\texttt{#1}}}
\newcommand{\many}[1]{\overline{#1}}
\newcommand{\Div}{\ |\ }

\newcommand{\critFn}{\raisebox{-.37em}{\resizebox*{1em}{!}{\includegraphics{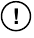}}}}
\newcommand{\genFn}{\raisebox{-.46em}{\resizebox*{1.2em}{!}{\includegraphics{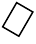}}}}
\newcommand{\cloudFn}{\raisebox{-.4em}{\resizebox*{1.5em}{!}{\includegraphics{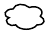}}}}
\newcommand{\tapp}{t\app}
\newcommand{\app}{\textsf{APP}\xspace}

%% file: intro.tex
% !TeX root = main.tex
\section{Introduction}
\label{sec:introduction}

%\todo[inline]{
%Introduce Serverless
%\\ Introduce Serverless function Scheduling
%\\ Briefly recap ICSOC 2020\cite{PGMZ20}
%\\ Recall the problem of cluster-aware scheduling in\app
%\\ Describe the contributions of this paper:
%\\\indent - an extension of the\app language to describe cluster-aware function scheduling
%\\\indent - prototype implementation of the extension in OpenWhisk
%\\\indent - definition of a suite for benchmarking serverless function scheduling 
%\\\indent - benchmarks of our prototype and comparison with some of the main Serverless platforms (Vanilla OpenWhisk, Fission, and OpenFaaS)
%}

%Il use case lo presenterei come risultato di collaborazione
%con una azienda manifatturiera interessata ad un utilizzo del 
%cloud in un continuo cloud-edge. Avevano interessa all'utilizzo
%del paradigma serverless per ridurre al minimo l'effort
%nella gestione dell'infrastruttura (ad esempio, deployment di specifici
%servizi nei vari punti del continuo cloud-edge).
%Poi dobbiamo essere convincenti nel dire che le attuali tecnologie
%serverless managed IoT-Cloud (Azure IoT, AWS Greengrass,..) non vanno bene! 
%Qui serve fare molta attenzione, e si deve essere convincenti!

%\todo{jac. Concordo con Gigio. Ho provato a fare una passata. La nostra soluzione dovremmo spostarla in sezione 3}

%\todo[inline]{Bisogna dire che assunzioni facciamo? Ci focalizziamo su OpenWhisk e assumiamo kubernetes? Poi dire che cmq il nostro approccio può essere proposto per altri sistemi.}

Serverless is a cloud service that lets users deploy architectures as compositions of stateless functions, delegating all system administration tasks to the serverless
platform~\cite{Jonas-etal:BerkeleyViewOnServerless}. %providers 
%the duty to manage their deployment and scaling. Hence, although a bit of a
%misnomer---as servers are of course involved--- The ``less'' in Serverless
%refers to the removal of some
%
This has two benefits for users. First, they save time by delegating resource
allocation, maintenance, and scaling to the platform. Second, they pay only for the resources that perform work avoiding the costs of running idle servers.
% In this way, the application developers are free from the management of
% server-related concerns like maintenance and scaling, as well as from the
% expenses deriving from their partial utilization like it happens, e.g., with
% idle servers.

For example, Amazon AWS Lambda,
% \footnote{\url{https://aws.amazon.com/lambda/}},
%~\cite{web:IntroducingAwsLambda},
Google Cloud
Functions,
% \footnote{\url{https://cloud.google.com/functions/}}, 
and Microsoft Azure
Functions\footnote{Resp. \url{https://aws.amazon.com/lambda/}, \url{https://cloud.google.com/functions/}, \url{https://azure.microsoft.com/}.} are managed serverless offers by popular
cloud providers, while OpenWhisk,
% \footnote{\url{https://openwhisk.apache.org/}},
OpenFaaS,
% \footnote{\url{https://www.openfaas.com/}},
OpenLambda, and
% ~\footnote{\url{https://github.com/open-lambda/open-lambda}}
% ~\cite{HSHVAA16}, and
Fission\footnote{Resp. \url{https://openwhisk.apache.org/}, \url{https://www.openfaas.com/}, \url{https://github.com/open-lambda/open-lambda}, \url{https://fission.io/}.} are open-source alternatives, used also in private
deployments.

In all these cases, the platform manages the allocation of function executions
over the available computing resources, also called \emph{workers}. However, not
all workers are equal when allocating functions. Indeed, effects like \emph{data
  locality}~\cite{HSHVAA16}---due to high
latencies to access data---or \emph{session
  locality}~\cite{HSHVAA16}---due to the
need to authenticate and open new sessions to interact with other services---can
sensibly increase the run time of functions. These issues become more prominent
when considered in multi-region (i.e., when the application uses cloud resources located in different regions)
% , multi-cloud (i.e., when the application is using one or more cloud provider)
and in the cloud-edge continuum (i.e., the cloud is encompassed also by edge nodes with computing, memory, or energy consumption constraints) where only
specific workers can access some local resources.

\paragraph{Tackling the Serverless Function Scheduling Problem}

\begin{figure}[t]
  \centering
  \includegraphics[width=\columnwidth]{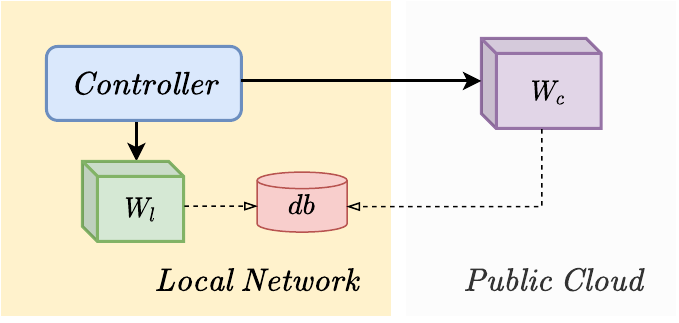}
  \caption{Example of function-execution scheduling problem.}
  \label{img:use-case_excerpt}
\end{figure} 
To visualise the problem, consider an excerpt of the running example used in
this article, reported in \cref{img:use-case_excerpt}. There, 
% , which depicts an excerpt of the use case we present in this paper (in
% \cref{sec:motivating_example}) to motivate our contribution. In the example we
% have
we have a simple serverless system composed of two workers. One worker, \(W_l\),
executes in the local network and the other, \(W_c\), is in a public cloud. Both
workers can execute functions that interact (represented by the dashed lines)
with a database {\it db} deployed in the local network. When the
\(\textit{Controller}\) (acting as function scheduler) receives a request to
execute the function, it must decide on which worker to execute it. To minimise
the response time, the \(\textit{Controller}\) should consider the different
computational power of the workers as well as their current loads, which
influence the time they take to execute the function. Moreover, for those
functions that interact with the database, also the latency to access the latter
plays an important role in determining the performance of function execution:
\(W_l\) is close to the {\it db} and enjoys a faster interaction with it while
\(W_c\) is farther from it and can undergo heavier latencies.
%\todo{jac: parliamo di site 1 e site 2? Non ha senso più dire italy or greece
%visto che non facciamo vedere gli esperimenti di icsoc. Forse non ha neanche
%senso mettere questo esempio quando abbiamo il caso più elaborato dopo}

\paragraph{Challenges towards Customisable Serverless Scheduling Policies}
The first step to achieving customisable serverless scheduling policies is
finding a way to \emph{express} the relationship between functions and workers.
For instance, we want to let the user indicate what workers the function shall
run upon and their priority, e.g., in \cref{img:use-case_excerpt}, the user
might want to run functions that heavily interact with \(\mathit{db}\) on
\(W_l\), to minimise latency. This item raises the initial challenges
(C1--2):

\begin{quote}
\em{C1. Express the relation between functions and the workers that can run
them.}
\end{quote}
\begin{quote}
\em{C2. Define the priority to select the worker that shall run a function instance.}
\end{quote}
%
%(e.g., depending on the state of the system). 
Our idea to tackle C1 is to assume the association of each worker with a
distinct label that univocally identifies it. Then, we need to allow users to
specify the collections of workers that can execute a given function. We enclose
these collections in a block, and we associate each with a distinct label. The
user can pair any function with the block label that describes its scheduling
logic. In this way, we know what block to consider when scheduling any given
function. For instance, in the example in \cref{img:use-case_excerpt}, we can
specify the preference between \(W_l\) and \(W_c\), which we interpret as worker
labels, with a block that we call \verb|db_query|, and we associate all the
functions which heavily interact with \(\mathit{db}\) with the \verb|db_query|
tag. To express the information on the labelled blocks, we can adopt a
declarative representation. We choose to represent this information in YAML
format~\cite{YAML}, which is a widely-adopted standard serialisation format for
DevOps tools like Kubernetes, Docker, etc. The code below shows the YAML
definition of the \verb|db_query| block.

\begin{lstlisting}[language=yaml, backgroundcolor=\color{Gold1!20},mathescape=true]
- db_query:
  - $\hl{workers}$: 
    - $\hl{wrk}$: $W_{l}$
    - $\hl{wrk}$: $W_{c}$
    $\hl{strategy}$: $\hlopt{best\_first}$
\end{lstlisting}

Above, we pair the block \verb|db_query| with the list of workers
\(W_l\) and \(W_c\) and use the keyword \hl{strategy} associated with the
\hlopt{best\_first} option to tackle challenge C2, i.e., we express the priority
of selection as a parameter of the block. In the example, \hlopt{best\_first}
indicates that we choose the workers depending on their order of appearance in
the list.

Although solving C1--2 allows us to describe the relationship between functions
and workers, we need to make the scheduling logic able to handle the elasticity
of Cloud resource management, i.e., that computing nodes can appear or disappear
at runtime. In our case, this means that new workers can enter the system and
existing workers can leave. This raises the challenge:

\begin{quote}
\em{C3. Express scheduling policies that refer to a dynamically modifiable set
of workers.}
\end{quote}

To deal with C3, we introduce the notion of \emph{worker set}s. Each worker set
is uniquely identified by a label. Workers can be associated with
multiple worker-set labels. Hence, while we can assume a fixed collection of
worker-set labels, the number of workers associated with a given worker-set
label can change at runtime (e.g., workers join/leave a worker set
at runtime).
% and we can associate the same worker with multiple worker-set
% labels.
% assume a statically defined collection of possible worker set labels, but new
% workers can be dynamically added to the system, and these can be labeled with
% one, or more, of such labels. 
For instance, in \cref{img:use-case_excerpt}, if we interpret \(W_l\) and
\(W_c\) not as worker labels, but as worker-set labels---where \(W_l\) labels
workers belonging in the set of \(\mathit{Local\ Network}\) and \(W_c\) labels
workers in the \(\mathit{Public\ Cloud}\)---we can modify the previous code
example to use dynamic sets of workers.

\begin{lstlisting}[language=yaml, backgroundcolor=\color{Gold1!20},mathescape=true]
- db_query:
  - $\hl{workers}$: 
    - $\hl{set}$: $W_{l}$
    - $\hl{set}$: $W_{c}$
    $\hl{strategy}$: $\hlopt{best\_first}$
\end{lstlisting}

In the snippet above, we indicate that $W_{l}$ and $W_{c}$ refer to worker
\hl{set}s. The syntactic change from the previous code excerpt is minimal, but
the effect on the scheduling logic is sensible. Now, the scheduler will try to
run a \verb|db_query|-tagged function on one of the nodes present in \(W_l\) (at
scheduling time) and the only way for the scheduler to run the function on one
of the workers in \(W_c\) is that none of the nodes in \(W_l\) can host the
function.
% hence before moving to one of the workers of the worker set $W_{c}$ in the
% cloud, it is necessary to check the availability of at least one worker of the
% set $W_{l}$ present in the local network.

Worker sets allow us to capture the elasticity of Cloud resource management;
however, they are not powerful enough to describe multi-region or cloud-edge
scenarios. For instance, in these settings, a serverless system could span
over different data centers, each with its characteristics: features of the
computing nodes, interaction mechanisms, availability of persistent storage,
elasticity (i.e., the possibility to add/remove computing nodes at runtime),
etc. Different computing environments
%, that we simply call \emph{zones}, 
% with different characteristics
% , 
could require different function scheduling
policies. Hence, we need to tackle the challenge:

\begin{quote}
\em{C4. Express scheduling policies that adapt to the different characteristics
of multi-region and cloud-edge scenarios.}
\end{quote}

To undertake C4, we introduce the notion of \emph{zone}. Each worker belongs in
one zone. Zones allow us to express \emph{topologies}, i.e., a partition of the
computing resources in distinct zones. For instance, in the example in
\cref{img:use-case_excerpt}, we can consider two zones, \(\mathit{Local\
Network}\) and \(\mathit{Public\ Network}\), and pair each of the workers in the
worker set $W_{l}$ with the former, and those in the set $W_{c}$ with the
latter. While zones let us model (through partitioning) a given
multi-region/cloud-edge system, we need a way to associate the same function
label to the different scheduling policies that characterise each, distinct
zone. This motivates the introduction of a new class of elements %in our
%declarative language: labelled 
called \hl{controller}s. Functionally, a controller is a component of serverless
platform architectures that coordinates the allocation of functions over a set
of workers.
%---where the controller-worker association is many-to-many. 
Here, we
assume to associate each %labelled 
controller with one zone (multiple controllers
can belong to the same zone for, e.g., load balancing purposes). 
%Controllers allow us to define topology constraints on where functions can run,
%abstracting the language from the physical location/zone-labelling of workers.
%To support the definition of different scheduling policies for the different
%zones, we need to enrich our systems with additional components. In fact,
%workers are usually mere executors of functions and are not responsible for the
%implementation. For this reason, we also consider \emph{controllers},
%components specifically dedicated to the implementation of the function
%scheduling policies, and we place our controllers over the system topology,
%i.e., each controller is paired with one zone. 
Thanks to controllers, we can differentiate function scheduling policies over
different zones. To do it, we associate a controller in a zone of interest with
the worker blocks and the worker selection mechanisms specific to both that zone
and the functions that shall run there. 

To exemplify the relationship between functions, controllers, and workers, we
slightly modify the architecture depicted in \cref{img:use-case_excerpt}. First,
we introduce a new controller, called \(\mathit{Controller\_cloud}\). Second, we
assume that \(\mathit{Controller}\) and \(W_l\) workers belong to the
\(\mathit{Local\ Network}\) zone, and \(\mathit{Controller\_cloud}\) and \(W_c\)
workers belong to the \(\mathit{Public\ Cloud}\) one. Third, we associate all
workers with the worker-set label \(\mathit{any}\). Lastly, we assume the
complete partitioning of resources in zones, such that the controller of a zone
can only allocate functions on workers in its same zone\footnote{In
\cref{contributions:sec:topology_invoker_distribution} we call this resource
partitioning policy \emph{isolated}.
% and it is such that each controller can only allocate functions on workers in its same zone. 
Other policies can exist and we discuss them in \cref{contributions:sec:topology_invoker_distribution}.}.

Now, we can extend our previous example to use topology-defined boundaries.
Specifically, we instruct the scheduler to first try to schedule
\verb|db_query|-tagged functions within the zone of \(\mathit{Controller}\). If
none of the workers in the \(\mathit{Local\ Network}\) is available, we defer
the execution to the \(\mathit{Public\ Cloud}\) workers, managed by
\(\mathit{Controller\_cloud}\). We express this logic with the script below
(notice that \hl{strategy} has the same scope as the block label, indicating
that we follow the \hlopt{best\_first} order-of-appearance logic to select among
the controllers in the block).

\begin{lstlisting}[language=yaml, backgroundcolor=\color{Gold1!20},mathescape=true]
- db_query:
  - $\hl{controller}$: $\mathit{Controller}$ 
    $\hl{workers}$:
      - $\hl{set}$: $\mathit{any}$
  - $\hl{controller}$: $\mathit{Controller\_cloud}$ 
    $\hl{workers}$:
      - $\hl{set}$: $\mathit{any}$
  $\hl{strategy}$: $\hlopt{best\_first}$
\end{lstlisting}

%\todo[inline]{S: Bisognerebbe dire cosa otteniamo da questa associazione col controller. Se no è uguale ad avere i worker set. Possiamo introdurre $\hl{topology-tolerance}$: \hlopt{same} e dire che se la funzione non viene fatta girare sui workers del blocco, può ancora girare, ma solo su workers di un altro controller che sono nella stessa zona di \(Controller\). E.g., questo possiamo volerlo per avere più flessibilità dei worker sets, ma comunque mantenere controllo su dove possono andare a finire le funzioni (e.g., se finiscono su worker ``sconosciuti'', sappiamo che comunque appartengono alla Local Network). Il discorso è uguale per \hl{set} invece che \hl{wrk}, ma probabilmente basta \hl{wrk} per l'esempio.} 

Challenges C1--4 are more theoretical and regard finding proper language
constructs to describe a wide gamut of scheduling policies. However, going towards implementation, we see two closing challenges.
First, we shall be able to implement the language semantics,
i.e.,
\begin{quote}
  \em{C5. Show that assembling a serverless platform able to run custom serverless policies is feasible.}
\end{quote}
Second (and final challenge), we need to show that running custom serverless
policies is beneficial, i.e.,
\begin{quote}
  \em{C6. Show that running custom serverless scheduling policies increases the performance of serverless functions.}
\end{quote}
This last challenge is a nuanced one. On the one hand, we need to show that
custom scheduling policies reduce function run times in scenarios that allow for
such customisations (e.g., in locality-bound settings). On the other hand, we
must show that custom scheduling policies do not introduce sensible overhead in
contexts where customisation is not required.

We tackle C5 by producing a prototype from a popular open-source serverless
platform, Apache OpenWhisk---both upgrading existing components responsible for
function scheduling and adding new ones---so that it supports parsing and
execution of our YAML-based customised scheduling policies. We undertake C6 by
showing both qualitative and quantitative evaluations between our prototype and
vanilla OpenWhisk. Qualitatively, we verify, through a case study, that our
prototype can capture functional scheduling requirements that would need a much
more complex and fragile setup if done with vanilla OpenWhisk. Quantitatively,
we report benchmarks of a comprehensive set of serverless functions, showing
that the overhead introduced by our prototype is negligible w.r.t. the
performance of vanilla OpenWhisk.

\begin{figure*}[t]
  \begin{center}
  \includegraphics[width=.8\textwidth]{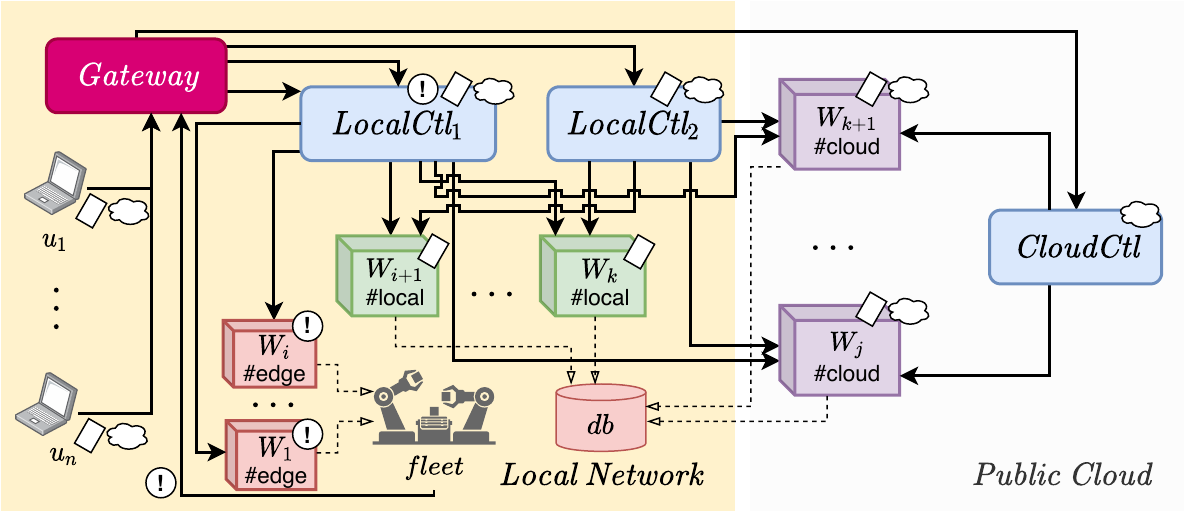}
  \end{center}
  \caption{Representation of the case study.}
  \label{fig:example}
\end{figure*}

%An example that motivates o
\paragraph{Motivating Example}
We further clarify the concepts above with a case study from a company
among our industry partners, which we use as an example throughout the article. We deem the case useful to help understand our
contribution and clarify the motivation behind our work.
%\todo{jac:secondo me basta questo esempio per spiegare tutto. Quello semplice
%icsoc non serve}

The case concerns a cloud-edge-continuum system to control and perform
both predictive maintenance and anomaly detection
over a fleet of robots in the production line. The system runs three categories
of computational tasks: i) predictions of critical events, performed by
analysing data produced by the robots, ii) non-critical predictions and generic
control activities, and iii) machine learning tasks.
% In particular, users periodically trigger machine-learning activities to
% improve the prediction algorithms by including new collected data. 
Tasks i) follow a closed-control loop between the fleet that generates data and
issues these tasks and the workers that run these tasks and can act on the
fleet. Since tasks i) can avert potential risks, they must execute with the
lowest latency and their control signals must reach the fleet urgently.
The users of the system launch the other categories of tasks. These are not
time-constrained, but tasks iii) have resource-heavy requirements.

We depict the solution that we have designed for the deployment of the system in
\cref{fig:example}. We consider three kinds of functions, one for each category
of tasks: critical functions~\critFn{} (in \cref{fig:example}), generic
functions~\genFn{}, and machine learning functions~\cloudFn{}.
To guarantee low-latency and the possibility to immediately act on the robots,
we execute \emph{critical} functions~\critFn{} on edge devices (workers
$W_{1},\ldots,W_i$ in \cref{fig:example}) directly connected to the robots.
Since machine-learning algorithms require a considerable amount of resources
that the company prefers to provision on-demand, we execute the machine-learning
functions \cloudFn{} on a public cloud, outside the company's perimeter
($W_{k+1},\ldots,W_j$ in \cref{fig:example}).
The generic functions~\genFn{} do not have specific, resource-heavy
requirements, but they might need to access the database {\it db} executing in the
local network. Hence, we schedule these preferably on the local cluster
($W_{i+1},\ldots,W_k$ in \cref{fig:example}) and use on-demand public-cloud
workers when the local ones are at full capacity.

For performance and reliability, our solution considers two function-scheduling
controllers for the internal workers, i.e., the controllers
$\mathit{LocalCtl_1}$ and $\mathit{LocalCtl_2}$, and one for cloud workers,
i.e., the controller $\mathit{CloudCtl}$. One local controller, namely
$\mathit{LocalCtl_1}$, has a dedicated low-latency connection with the edge
devices able to act on the fleet.

Finally, a $\mathit{Gateway}$ acts as load balancer among the controllers.
However, to follow the requirements of the company, instead of adopting a
generic round-robin policy, we need to instruct the $\mathit{Gateway}$ to
forward critical functions~\critFn{} to $\mathit{LocalCtl_1}$, the generic
functions~\genFn{} to one between $\mathit{LocalCtl_1}$ and
$\mathit{LocalCtl_2}$, and the cloud functions~\cloudFn{} to $\mathit{CloudCtl}$
(or to any other controller when the latter is not available).

%The main controller can schedule jobs over all the workers that are tagged with the tag $\mathit{\#internal}$ (i.e., $W_1,\ldots,W_i$) or $\mathit{\#cloud}$ while instead $\mathit{SecondC}$ can control only the workers tagged with $\mathit{\#non-critical}$ or $\mathit{\#cloud}$. The controller $\mathit{MainC}$ is the only one allowed to schedule the execution of
%critical functions while generic functions can be scheduled preferably by $\mathit{SecondC}$ and if this is not possible by $\mathit{MainC}$. The company also deploys an external controller 
%$\mathit{CloudC}$ to schedule the functions in the cloud.
%Finally, the entry point to the entire system is a $\mathit{Nginx}$ load balancer that manages the routing
%of function invocations to the appropriate controller.
%\todo{metterei un tag internal per tutti da W1 a Wi e poi un tag critical  su W1 e non-critical su W2,..,Wi). Così con i tag possiamo definire quello che vogliamo}

\newcommand{\code}[1]{\lstinline[language=yaml]{#1}}
\newcommand{\codeV}[1]{{\YAMLvaluestyle\texttt{#1}}}

%\todo[inline]{se il cambio di nomi va bene propagare le modifiche}
%
%This is an example where more controllers are used to share the computation resources without creating a partition of the resources. Naively, a possible solution to guarantee the scheduling of the functions as required would be to either i) creating 3 clusters, each with one controller or ii) create a cluster with one controller but partition the resources and detail the scheduling using approaches such as\app~\cite{PGMZ20} or ....
%Clearly, both these solutions have drawbacks: the first will not allow sharing of the resources while the second worsen the robustness and security of the system by requiring only one entry point to the system and requiring all workers to be shared.

%
% example of language for worker's deployment
% critical_workers:
%  main: \mathit{Controller}_1
%  policy: isolated

% internal:
%  main: \mathit{Controller}_1
%  others:
%   - \mathit{Controller}_2
%  policy: shared

% cloud_worker:
%  main: \mathit{Controller}_3
%  others:
%   - \mathit{Controller}_2
%   - \mathit{Controller}_1
%  policy: min_memory
%

\paragraph{Contributions}
The case above presents a scenario where we need to deploy the serverless
platform over at least a couple of zones (local network and public cloud) and
where the function-execution scheduling policy depends on a topology of
different clusters (edge-devices, local cluster, and cloud cluster). The
scheduling policies influence the behaviour of both the gateway and the
controllers, which need to know the current status of the workers (e.g., to execute generic functions in the cloud when the local cluster is overloaded).

One can obtain a deployment of the case by modifying the source code of all the
involved components and by hard-coding their desired behaviour. However,
this solution requires a deep knowledge of the internals of the components
and is a fragile solution, difficult to maintain and evolve.

The approach that we propose, presented in \cref{sec:tapp},
is based on
a new declarative language, called \tapp (Topology-aware Allocation Priority
Policies), used to write \emph{configuration} files describing topology-aware
function-execution scheduling policies.
% Jac: Commento la future reference. Non penso che serva For example, Figure
% \ref{fig:app_example} reports the configuration file for our motivating
% example. Given a configuration file, the components of the system adapts their
% behaviour to the specified policies. 
In this way, following the Infrastructure-as-Code philosophy, users (typically
DevOps) can keep all relevant scheduling information in a single repository (in
one or more \tapp files) which they can version, change, and run without
incurring downtimes due to system restarts to load new configurations.

After having presented \tapp, in \cref{sec:tapp_openwhisk}, we validate our
approach by implementing a serverless platform that supports \tapp-specified
scheduling policies, as an extension of OpenWhisk where worker selection happens
via tags associated to functions.
The implementation entailed the creation and inclusion in the existing
architecture of OpenWhisk of new components---e.g., a \emph{watcher} service,
which informs the gateway and the controllers on the current status of the nodes
of the platform---and the extension of existing ones with new
functionalities---e.g., to capture topological information at the level of
workers and controllers, to enable live-reloading of \tapp policies, to let
controllers and gateways follow \tapp policies depending on topological zones,
etc.\@.
% Finally, we also introduced the possibility to dynamically modify the set of
% available workers (thus including horizontal scalability) as well as the
% possibility to modify at run time the scheduling policy (thus supporting
% dynamic adaptation of the system). The architecture of our modified version of
% OpenWhisk is reported in Figure \ref{fig:openwhisk}. Jac: commento la prima
% parte che deve essere un minimal requirement. Commento la seconda visto che
% non è bellissimo cosa facciamo visto che poi bisogna mandare un comando per
% aggiornare i controller

Beyond proving our idea realisable, our \tapp-based OpenWhisk version allows us
to evaluate the feasibility and performance of topology-aware scheduling
policies. To do that, in \cref{sec:testing}, we define a case study to
qualitatively evaluate support for topology awareness in \tapp, and we devise a
set of deployments of the platform and several use cases that we use to compare
our version against vanilla OpenWhisk. Qualitatively, we show that our prototype
can capture functional scheduling requirements that would need a much more
complex and fragile setup if done with vanilla OpenWhisk. Quantitatively, in all
presented cases, our version is on par with or outperforms vanilla OpenWhisk.
\todo{qui c'è una ripetizione di quanto diciamo prima in C5 e C6. Secondo me 
dovremmo evitare la ripetizione. A occhio si può semplificare questa parte. 
GIGIO: secondo me questa ripetizione non e' grave, siamo in una sezione contribution,
ed in qualche modo e' giusto enfatizzare anche la parte validation tra le contribution,
e la ripetizione e' di una dozzina di righe.. poco per un paper di una ventina di pagine.}

This article integrates and extends material from~\cite{PGMZ20,DGMTZ22}. The
first introduces a prototypical version of \tapp, called \app, and proposes the
tag-to-function mechanism, and implements it in OpenWhisk. The second extends
\app to handle topological information and integrates new components and
functionalities in the OpenWhisk architecture to support \tapp-based policies.
The main new material introduced by this article regards the empirical
analysis reported in \cref{sec:testing}, which comprises both qualitative and quantitative evaluations. We also provide a revised
and extended presentation of both the language (\cref{sec:tapp}) and of the relevant technical details of our OpenWhisk \tapp-based extension and of the Infrastructure-as-Code solutions we offer to help users automatise the deployment of our extension (\cref{sec:tapp_openwhisk}).

%\paragraph{Structure of the paper}
%We structure the remainder of the paper as follows. In \cref{sec:preliminaries}
%we introduce preliminaries on the architecture of the OpenWhisk serverless
%platform.
%% and of Kubernetes, the container-based deployment technique that we have
%% adopted in the implementation of our modified version of OpenWhisk. 
%We present our main contributions in
%\cref{sec:implementation,sec:capp_openwhisk}, where we respectively detail our
%declarative language for expressing cluster-aware function-execution scheduling
%policies (\tapp) and we report on our implementation of a \tapp-based OpenWhisk
%platform.
%%
%We evaluate \tapp and our implementation in \cref{sec:testing,sec:results}, where
%we describe and comment on the methodology we followed to select our set of
%tests (either created ad-hoc or drawn from the literature), the deployment we
%used to run them, and our experimental results.
%%
%Finally, we close by comparing with related work in \cref{sec:related} and draw
%concluding remarks and future directions in \cref{sec:conclusion}.
%

%% file: preliminaries.tex
% !TeX root = main.tex
\section{Preliminaries}
\label{sec:preliminaries}

Before detailing our solution, we give some preliminary information useful to
understand its motivations and technicalities. First, we outline the problems
that motivate our research---as found in the literature. Then, we give an
overview of the OpenWhisk Serverless platform, which we use to
implement a prototype of our solution to the function scheduling problem.

\begin{figure*}[t]
  \begin{center}
    \includegraphics[width=.8\textwidth]{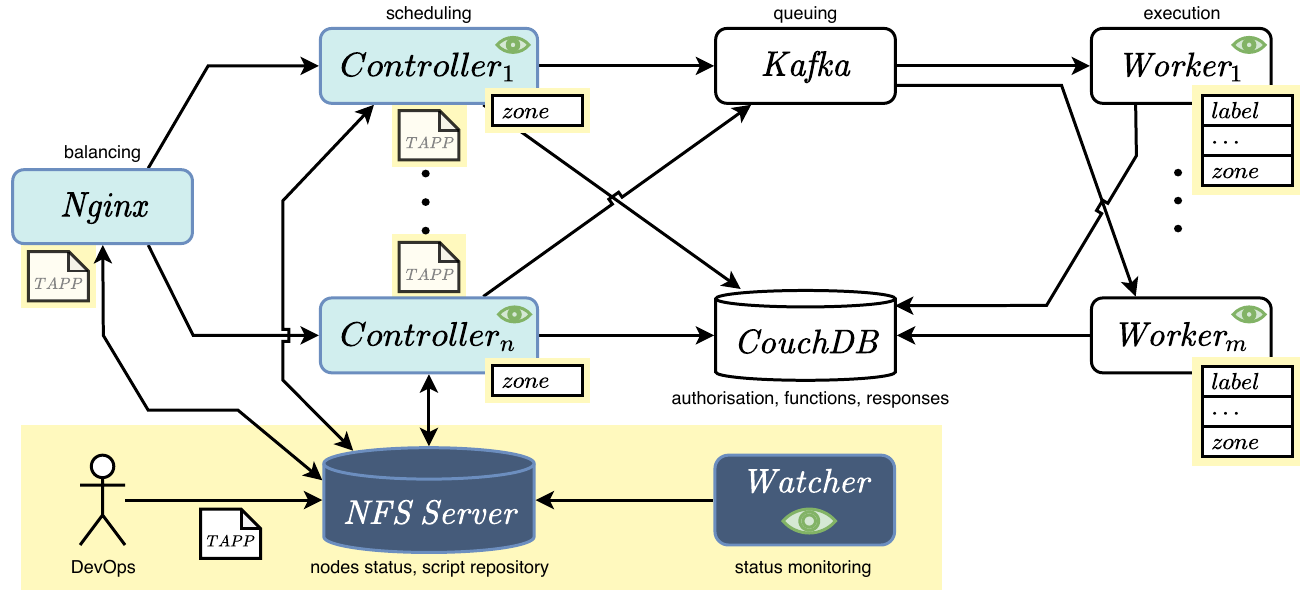}
  \end{center}
  \caption{Architectural view of our OpenWhisk extension. We highlight in \colorbox{LightCyan2}{light blue} the existing components of OpenWhisk we modified and in {\colorbox{yellow!30}{yellow}} the new ones we introduced.}
  \label{fig:openwhisk}
\end{figure*}

\paragraph{Serverless Function Scheduling}
The Serverless development cycle is divided in two main parts: \emph{a}) the
writing of a function using a programming language supported by the platform
(e.g. JavaScript, Python, C\#) and \emph{b}) the definition of an event that
should trigger the execution of the function. For example, an event is a request
to store some data, which triggers a process managing the selection,
instantiation, scaling, deployment, fault tolerance, monitoring, and logging of
the functions linked to that event. A Serverless provider---like IBM Cloud
Functions~\cite{web:ibm_functions} (using Apache OpenWhisk~\cite{web:OpenWhisk}), AWS
Lambda~\cite{web:aws_lambda}, Google Cloud Functions~\cite{web:googlefunctions}
or Microsoft Azure Functions~\cite{web:azurefunctions}---is responsible
to schedule functions on its workers, to control the scaling of the
infrastructure by adjusting their available resources, and to
bill its users on a per-execution basis.

When instantiating a function, the provider has to create the appropriate
execution environment for the function. Containers~\cite{B14} and Virtual
Machines~\cite{FetAl09} are the main technologies used to implement isolated
execution environments for functions. How the provider implements the allocation
of resources and the instantiation of execution environments impacts on the
performance of the function execution. If the provider allocates a new container
for every request, the initialisation overhead of the container would negatively
affect both the performance of the single function and heavily increase the load
on the worker.
A solution to tackle this problem is to maintain a ``warm'' pool of
already-allocated containers. This matter is usually referred to as \emph{code
locality}~\cite{HSHVAA16}.
Resource allocation also includes I/O operations that need to be properly
considered. For example, the authors of~\cite{wang2018peeking} report that a
single function in the Amazon serverless platform can achieve on average 538Mbps
network bandwidth, an order of magnitude slower than single modern hard drives
(the authors report similar results from Google and Azure). Those performance
result from bad allocations over I/O-bound devices, which can be reduced
following the principle of \emph{session locality}~\cite{HSHVAA16}, i.e., taking
advantage of already established user connections to workers.
Another important aspect to consider to schedule functions, as underlined by the
example in \cref{img:use-case_excerpt}, is that of \emph{data locality}, which
comes into play when functions need to intensively access (connection- or
payload-wise) some data storage (e.g., databases or message queues).
Intuitively, a function that needs to access some data storage and that runs on
a worker with high-latency access to that storage (e.g., due to physical
distance or thin bandwidth) is more likely to undergo heavier latencies than if
run on a worker ``closer'' to it. Data locality has been subject of research in
neighbouring Cloud contexts~\cite{xie2016pandas,wang2014maptask}.

%In this section, we introduce the main concepts and components of serverless platforms. We detail in particular Apache OpenWhisk, which we extend in this work.

\paragraph{Apache OpenWhisk}
% \paragraph{Serverless Platforms}
The main proprietary serverless platforms are AWS Lambda, Google Cloud
Functions, and Azure Functions.
% ~\cite{HBS21}. 
Despite their large user base,
these solutions do not provide source code one could build upon and offer
limited documentation on their inner workings. Fortunately, customisable and
evolvable open-source alternatives are gaining traction in the serverless
market. The most popular ones are Apache OpenWhisk, OpenFaaS, and
OpenLambda. %~\cite{HBS21}.
% As expected, different platforms have their unique architectures and
% technology stacks, resulting in different approaches to function scheduling
% and deployment.
%
Since these platforms have similar architectures~\cite{HBS21}
%  (see \cref{sec:related} for a
% brief comparison) 
we describe their main components via
the most popular, open-source one: Apache OpenWhisk. %~\cite{HBS21}.

% \paragraph{Apache OpenWhisk}

Apache OpenWhisk is an open-source, %production-ready
serverless platform initially developed by IBM and donated to the
Apache Software Foundation.
We report in \cref{fig:openwhisk} a scheme of the architecture of OpenWhisk. For
compactness, we include in \cref{fig:openwhisk} the modified and new elements
introduced by our extension, which we describe in \cref{sec:tapp_openwhisk}.
Here, we focus only on the original components and functionalities of OpenWhisk.

In \cref{fig:openwhisk}, from left to right, %of \cref{fig:openwhisk}, 
we first find \textit{Nginx}, which acts as the gateway and load balancer to
distribute the incoming requests. \textit{Nginx} forwards each request to one of
the \textit{Controller}s in the current deployment.

The \textit{Controller}s are the components that decide on which of the
available computation nodes, called \textit{Worker}s\footnote{OpenWhisk's documentation uses the more specific term ``invokers''.
%   Here, we use the more abstract term ``workers''.
  %Since we use OpenWhisk but do not depend on it, we adopt the generic term ``workers''.
  }, to schedule the execution of a given function.
\textit{Controller}s and \textit{Worker}s do not interact directly but use
Apache \textit{Kafka}~\cite{KNR11} and \textit{CouchDB}~\cite{ALS10} to
respectively handle the routing and queueing of execution requests and 
to manage the authorisations and the storage of functions and 
of their outputs/responses.

\textit{Worker}s execute functions using Docker containers. To schedule
executions, \textit{Controller}s follow a hard-coded policy that mediates load
balancing and caching. This works by trying to allocate requests to the same
functions on the same %(pool of) 
\textit{Worker}s\footnote{More precisely, the OpenWhisk allocation policy, called
  ``co-prime scheduling'', associates a function to a hash and a step size. The hash finds the primary worker. The step size finds a list of workers used in succession when the preceding ones become overloaded.
  % ---a number that is co-prime of and smaller than the number of
  % workers. The identifier of the primary worker results from the modulo between
  % the hash and the number of workers. In case the primary is invalid, we find the
  % secondary workers by increasing the function hash with the step-size and
  % repeating the modulo operation.
}, hence saving time by skipping the
retrieval of the function from \textit{CouchDB} and the instantiation of the
container %by using a version 
already cached in the memory of the
\textit{Worker}.

%% file: implementation.tex
% !TeX root = main.tex
\section{Topology-aware Serverless Scheduling} %(tAPP)}
\label{sec:tapp}

We can now introduce the \tapp language. We do this first by means of a simple example that we use to introduce the structure of \tapp scripts. Then, we move on, presenting the principles behind the inception of \tapp, its syntax and semantics, closing with the definition of a possible \tapp script that captures the case study presented in \cref{sec:introduction}.

\input{img/app_syntax}

\subsection{Introduction of \tapp, by example}

The Topology-aware Allocation Priority Policy (\tapp) language is a declarative
language able to specify customised load-balancing policies and overcome the
inflexibility of the hard-coded load-balancing ones. The idea is that \tapp can
support developers and providers in optimising the execution of serverless
functions. \tapp is tailored to adapt to the different types of information on
the serverless infrastructure that providers share with developers. For example,
in edge deployments (where it is important to know on which machine functions run), developers know which nodes are available and their position; in the
Cloud, developers think in terms of regions (e.g., west/east US, Europe) and
zones (Los Angeles, New York, Paris areas), rather than single nodes. As
exemplified in \cref{sec:introduction}, \tapp policies can scale according to
these needs and adapt to edge-cloud-continuum scenarios, where policies can span
single nodes, unbound collections of these (e.g., defined by some common trait), and topological zones. \tapp can also work in the absence of
information provided to developers---without function-specific configurations,
\tapp-based platforms follow a default strategy, like the other, hardwired
alternatives.
  
As an extension of the example depicted in \cref{img:use-case_excerpt}, consider
some functions that need to access a database. To reduce latency (as per data
locality principle), the best option would be to run those functions on the same
pool of machines %/workers 
that run the database. If that option is not valid,
then running those functions on workers in the proximity (e.g., in the same
network domain) is preferable than using workers located further away (e.g., in
other networks).
We comment below an initial \tapp script that specifies the scheduling policies
only for those workers belonging to the pool of machines running the
database.

\begin{lstlisting}[language=yaml, backgroundcolor=\color{Gold1!20},mathescape=true]
- couchdb_query:
  - $\hl{workers}$: 
    - $\hl{wrk}$: DB_worker1
    - $\hl{wrk}$: DB_worker2
   $\hl{strategy}$: $\hlopt{random}$
   $\hl{invalidate}\hlstr:$ $\hlopt{capacity\_used}$ $\hlopt{50\%}$
  $\hl{followup}$: $\hlopt{fail}$
\end{lstlisting}

At the first line, we define the \emph{policy tag}, which is
\verb|couchdb_query|. As explained below, %in more depth 
%in the continuation of the section, 
tags are used to link policies to functions. Then, the keyword \hl{workers}
indicates a list of \emph{worker} (\hl{wrk}) labels, which identify the workers
in the proximity of the database, i.e., \verb|DB_worker1| and \verb|DB_worker2|.
As explained below, %in more depth 
%in the continuation of the section, 
labels are used to identify workers. Finally, we define three parameters: the
\hl{strategy} used by the scheduler to choose among the listed worker labels,
%\hl{workers}, 
the policy that \hl{invalidate}s the selection
%usage
of a %selected 
worker label, and the \hl{followup} policy in case all workers are
\hl{invalidate}d. In the example, we select one of the two labels
\hlopt{random}ly, we \hl{invalidate} their usage if the workers corresponding to
the chosen label are used at more than the \hlopt{50\%} of their capacity
(\hlopt{capacity\_used}) and, in case all workers are invalidated
(\hl{followup}), we let the request for function execution \hlopt{fail}.

\subsection{The \tapp approach}

Essentially, \tapp relies on \emph{policy tag}s that associate functions to
scheduling policies. A tag identifies a policy (e.g., we can use a tag
``critical'' to identify the scheduling behaviour of the critical~\critFn{}
functions of our case study, cf. \cref{sec:introduction}) and it marks all those
functions that shall follow the same scheduling behaviour (e.g., marking as ``critical'' any function that falls into that category).

Topologies are part of policies and come in two facets.
	{\em Physical} topologies relate to {\em zones}, which can represent
availability zones in public clouds and plants in multi-plant industrial
settings.
	% \todo{Jac: Secondo me non occorre dire tanto di più di zone e multiplant. Eliminerei la parte seguente fino a Locical topologies.}
	% Physical topologies capture the concerns of the \emph{platform
	% administrators}\footnote{Concerns include limiting resource grabbing and
	% shielding sensible hardware, depending on the deployment determined by the
	% DevOps (see \cref{contributions:sec:topology_invoker_distribution})}, who
	% decide how local and remote controllers can access workers (e.g., in our case
	% study, the \(CloudCtl\) cannot access workers in the local network).
	%
	{\em Logical} topologies instead represent partitions of workers.
The logical layer expresses the constraints of the \emph{user} and identifies
the pool of workers which can execute a given function (e.g., for performance).
The smallest logical topology is the singleton, i.e., a worker, which we
identify with a distinct label (e.g., \(W_1\) in \cref{fig:example}). In
general, policies can target lists of singletons as well as aggregate
multiple workers in different sets.

The interplay between the two topological layers determines which workers a
controller can use to schedule a function. For example, we can capture the
scheduling behaviour of the critical functions of our case study in this way: 1)
we assign \(\mathit{LocalCtl_1}\), \(\mathit{LocalCtl_2}\), and
\(W_1,\ldots,W_k\) to the same zone, 2) we configure said workers to only accept
requests from co-located controllers (this, e.g., excludes access to
\(\mathit{CloudCtl}\)), and 3) we set the policy of the critical functions to
only use the workers tagged with the \textit{edge} label, \textsf{\#edge} in
\cref{fig:example}.

Besides expressing topological constraints, policies can include other
directions such as the strategy followed by the controller to choose a worker
within the pool of the available ones (e.g., to balance the load evenly among
them) and when workers are ineligible (e.g., due to their resource quotas).

\subsection{The \tapp language}
\label{sec:tapp_language}

We report the syntax of \tapp in \cref{fig:capp_syntax}.
% \todo{specifichiamo meglio come sono gli identifiers? Tipo Identifiers ::= [a-zA-Z0..9]* ed eventualmente caratteri speciali? Il policy tag lo possiamo anche scrivere con EBNF}

% \todo{perchè abbiamo *(label) per i set se ora abbiamo la keyword set? Non possiamo mettere set label ... | all ...? 
% Faccio notare che non possiamo usare * in YAML perchè ha un altro significato. Cambiamo * con una altra keyword come all o altro?}

%
%We report in \cref{fig:capp_syntax} the syntax of\app: the
%\newFrag{highlighted} fragments are part of our extension; the others
%correspond to the original syntax of\app.
%
% In the following, we first briefly describe the original components of an\app
% script and their effect of the scheduling of serverless functions. Then, we
% detail the new syntactic elements introduced in this work and how these expand
% the expressiveness of\app scripts to describe cluster-aware priority policies.
% Finally, we further illustrate the language by reporting and commenting on a
% possible \app script that captures the requirements of \cref{fig:example}.
%
\tapp scripts are YAML~\cite{YAML} files. The basic entities considered in the language are \emph{a}) scheduling policies,
defined by a \textit{policy tag} identifier to which users can associate their
functions---the policy-function association is a one-to-many relation---and
\emph{b}) workers, identified by a \textit{worker label}---where a label
identifies a collection of computation nodes. All identifiers are strings formed with the accepted character set 
as defined in \cite{YAML}.

Given a tag, the corresponding policy
includes a list of blocks, possibly closed with \hl{strategy} and \hl{followup}
options. A block includes four parameters: an optional \hl{controller} selector,
a collection of \hl{workers}, a possible scheduling \hl{strategy}, and an
\hl{invalidate} condition. The outer \hl{strategy} defines the policy we must
follow to select among the blocks of the tag, while the inner \hl{strategy}
defines how to select workers from the items specified within a chosen
\hl{workers} block. The \hl{controller} defines the identifier of a specific
controller we want the gateway to redirect the invocation request to. When used,
it is possible to define a \hl{topology\_tolerance} option to further refine how
\tapp handles failures (of controllers). The collection of \hl{worker}s can be
either a list of labels pointing to specific workers (\hl{wrk}), or a worker
\hl{set}. In lists, the user can specify
the \hl{invalidate} condition of each single worker, while in sets, the
\hl{invalidate} condition applies to all the workers included in the set. When
users specify an \hl{invalidate} condition at block level, this is directly
applied to all \hl{workers} items (\hl{wrk} and \hl{set}) that do not define
one. In \hl{set}s the user can also specify a \hl{strategy} followed to choose
workers within the set. Finally, the \hl{followup} value defines the behaviour
to take in case no specified controller or worker in a tag is available to
handle the invocation request.
% \todo{ATTENZIONE: questa descrizione e' vecchia, e non include le informazioni
% topologiche. Va bene fare una presentazione incrementale, prima senza
% topologia e poi con, ma bisogna spiegerlo}

We discuss the \tapp semantics, and the possible parameters, by commenting on a
more elaborate script extending the previous one, shown in
\cref{fig:APPscript}. The \tapp script starts with the tag \verb+default+, which
is a special tag used to specify the policy for non-tagged functions, or to be
adopted when a tagged policy has all its members invalidated, and the
\hl{followup} option is \hlopt{default}.

In \cref{fig:APPscript}, the \verb+default+ tag describes the default behaviour
of the serverless platform running \tapp. In this case we use a \hl{workers} \hl{set} to select workers,
with no value specified for \hl{set} which represents all worker labels. The \hl{strategy} selected is
the \hlopt{platform} default. In our prototype in \cref{sec:tapp_openwhisk} the
\hl{platform} strategy corresponds to a selection algorithm, hinted in
\cref{sec:preliminaries}, which mediates load balancing and code locality by
associating a function to a numeric hash and a step-size---a number that is
co-prime of and smaller than the number of workers\footnote{The identifier of
the primary worker results from the modulo between the hash and the number of
workers. In case the primary is invalid, we find the secondary workers by
increasing the function hash with the step-size and repeating the modulo
operation.}. The \hl{invalidate} strategy considers a worker non-usable
when it is \hlopt{overload}ed, i.e., it does not have enough resources to run
the function.

\begin{figure}[t]
	\begin{minipage}{\columnwidth}
	\begin{lstlisting}[language=yaml, backgroundcolor=\color{Gold1!20},mathescape=true]
	- default: 
			- $\hl{workers}$:
					- $\hl{set}$:
							$\hl{strategy}$: $\hlopt{platform}$
							$\hl{invalidate}$: $\hlopt{overload}$
	
	- couchdb_query:
			- $\hl{workers}$: 
					- $\hl{wrk}$: DB_worker1 
					- $\hl{wrk}$: DB_worker2
					$\hl{strategy}$: $\hlopt{random}$
					$\hl{invalidate}$: $\hlopt{capacity\_used}$ $\hlopt{50\%}$
			- $\hl{workers}$:
					- $\hl{wrk}$ : near_DB_worker1
					- $\hl{wrk}$ : near_DB_worker2
					$\hl{strategy}$: $\hlopt{best\_first}$
					$\hl{invalidate}$: $\hlopt{max\_concurrent\_invocations}$ $\hlopt{100}$
			$\hl{followup}$: $\hlopt{fail}$
	\end{lstlisting}
	\end{minipage}
	\caption{\label{fig:APPscript}Example of a \tapp script.}
	\end{figure}

Besides the \verb|default| tag, the \verb+couchdb_query+ tag is used for those
functions that access the database.
The scheduler considers worker blocks in order of appearance from top to bottom.
As mentioned above, in the first block (associated to \verb|DB_worker1| and
\verb|DB_worker2|) the scheduler randomly picks one of the two worker labels and
considers the corresponding worker invalid when it reaches the \hlopt{50\%} of
capacity. Here the notion of capacity depends on the implementation (e.g., our
OpenWhisk-based \tapp implementation in \cref{sec:tapp_openwhisk} uses
information on the CPU usage to determine the load of invokers). When both
worker labels are invalid, the scheduler goes to the next \hl{workers} block,
with \verb|near_DB_worker1| and \verb|near_DB_worker2|, chosen following a
\hlopt{best\_first} strategy---where the scheduler considers the ordering of the
list of \hl{workers}, sending invocations to the first until it becomes invalid,
to then pass to the next ones in order. The \hl{invalidate} strategy of the
block (applied to the single \hl{wrk}) regards the maximal number of concurrent
invocations over the labelled worker---\hlopt{max\_concurrent\_invocations},
which is set to \hlopt{100}. If all the worker labels are invalid, the scheduler
applies the \hl{followup} behaviour, which is to \hlopt{fail}. 

Users can define subsets of workers by 
specifying a label associated with the workers,
%  restricting the scope of the universal
% operator \hlopt{*} % Indeed, similarly to what described for topology zones,
% users can exploit existing custom labels of the workers for scheduling
% functions. Then, users can select subsets of the workers 
% suffixing it with a label, 
% e.g., \hlopt{*local} selects
e.g., \hlopt{local} selects
only those workers associated to the \textit{local} label.

The scheduling on
% \hlopt{*}-induced 
worker-\hl{set}s follows the same logic of
block-level worker selection: it
% presented in Algorithm \ref{alg:app_scheduling} (e.g., 
exhausts all workers before deeming the item invalid. Since worker-set
selection/invalidation policies are distinct from block-level ones, we let users
define the \hl{strategy} and \hl{invalidate} policies to select the worker in
the set. For example, we can 
pair the above selection with a \textit{strategy} and an \textit{invalidate}
options, e.g.,

\begin{lstlisting}[language=yaml, backgroundcolor=\color{Gold1!20},mathescape=true]
- workers: 
  - $\hl{set}$: $\hlopt{local}$
    strategy: $\hlopt{random}$
    invalidate: $\hlopt{capacity\_used}$ $\hlopt{50\%}$
\end{lstlisting}
which tells the scheduler to adopt the \hlopt{random} selection strategy and the
\hlopt{capacity\_used} invalidation policy when selecting the workers in the
\textit{local} set. When worker-sets omit the definition of the selection \hl{strategy} we consider the default one. When the invalidation option is omitted,
we either apply that of the enclosing block or, if that is also missing, the default one.

Summarising, given a policy tag, the scheduler follows the policy defined in the \hl{strategy} option to select the corresponding
blocks. A block includes three parameters:
  
\begin{itemize}
  \item \hl{workers}: which either contains a non-empty list of worker
  (\hl{wrk}) labels, each paired with an optional invalidation condition, or a
  worker-\hl{set} label (possibly blank, to select all workers) to range over sets of workers; workers \hl{set}s optionally define the \hl{strategy} and
  \hl{invalidate} options to select workers within the set and declare them invalid;

  \item \hl{strategy}: defines the policy of item selection at the levels of \textit{policy\_tag}, \textit{workers} block, and workers \hl{set}s.
 \app currently supports three strategies:

  \begin{itemize}
      
    \item \hlopt{random}: selects items in a fair random manner;
      
    \item \hlopt{best\_first}: selects items following their
    order of appearance;
      
    \item \hlopt{platform}: selects items following the default strategy
      of the serverless platform---in our prototype, this corresponds to a co-prime-based selection.
  
    \end{itemize}

  \item \hl{invalidate}: specifies when a worker (label) cannot host the
  execution of a function. All invalidate options include, as preliminary
  condition, the unreachability of a worker. When all labels in a block are
  invalid, we follow the defined \hl{strategy} to select the next block one until we either find a valid worker or we exhaust all blocks. In the latter case, we apply the \hl{followup} behaviour. Current
  \hl{invalidate} options are:

  \begin{itemize}
    
    \item \hlopt{overload}: the worker lack enough computational resources to
    run the function;\footnote{The kind of computational resources that
    determine the \hlopt{overload} option depends on the APIs provided by a
    given serverless platform. For example, in our prototype in
    \cref{sec:tapp_openwhisk} we consider a worker label \hlopt{overload}ed when
    the related invokers are declared ``unhealthy'' by the OpenWhisk APIs, which
    use memory consumption and CPU load.}

    \item \hlopt{capacity\_used}: the worker reached a threshold percentage of
    CPU load;

    \item \hlopt{max\_concurrent\_invocations}: the worker have reached a
    threshold number of buffered concurrent invocations.

  \end{itemize}
  
  \item \hl{followup}: specifies the policy applied when all the blocks in a
  policy tag are considered invalid. The supported followup strategies are:
  
  \begin{itemize}
    \item \hlopt{fail}: drop the scheduling of the function;
    \item \hlopt{default}: apply the \texttt{default} tag.
  \end{itemize}
  
\end{itemize}

Since the \texttt{default} block is the only possible ``backup'' tag used when all workers of a custom tag cannot execute a function (because they are all invalid), the \hl{followup} value of the \texttt{default} tag is always set to \hlopt{fail}.

% \todo{YAML invalido, non si possono mischiare liste e singoli valori}
\input{img/tapp_example}

Besides the above elements, to further detail topological constraints of
function execution scheduling, we have the \newFrag{\textit{controller}}.
This is an optional, block-level parameter that identifies which of the
possible, available \hl{controller}s in the current deployment we want to target
to execute the scheduling policy of the current tag. Similarly to workers, we
identify controllers with a label.
% \todo{con questo paragrafo parte l'aspetto topologico, ma viene introdotto
% senza alcun esempio mentre prima c'' un esempio per la parte non topologica. Questo rende asimmetrica la presentazione.}

% Note
% that, in \cref{fig:capp_syntax}.

As mentioned above, a \textit{controller} clause can have \hl{topology\_tolerance} as optional parameter.
When deploying controllers and workers, users can label them with the
topological \emph{zone} they belong in\footnote{Zone labels of controllers and
	workers are not used in %invisible to 
	\tapp scripts, 
	which only specify co-location constraints, i.e., requests to
	consider workers in the same zone of a given controller. 
	Zone labels are used by 
	the infrastructure to implement the \tapp constraints.}.
%	on their
%	relationship.}. 
Hence, when the designated \hl{controller} is unavailable, \tapp
can use this topological information to try to satisfy the scheduling request by
forwarding it to some alternative controller.

The \hl{topology\_tolerance} parameter specifies what workers an alternative
controller can use. Specifically,
% \begin{itemize}
% \item 
\hlopt{all} is the default and most permissive option and imposes no
restriction on the topology zone of workers;
% \item 
\hlopt{same} constrains the function to run on workers in the same
zone of the faulty controller (e.g., for data locality);
% \item
\hlopt{none} forbids the forward to other controllers.
% \end{itemize}
As an example, we could take advantage of the topology zones and rewrite the previous \tapp script from \cref{fig:APPscript} 
for the \verb+couchdb_query+ tag. e.g.,
\begin{lstlisting}[language=yaml, backgroundcolor=\color{Gold1!20},mathescape=true]
- couchdb_query:
  - controller: ~DBZoneCtl~
    workers: 
				 - $\hl{set}$: $\hlopt{local}$
				   strategy: $\hlopt{random}$
			 topology_tolerance: ~same~
  followup: ~default~
\end{lstlisting}
this way it is guaranteed that the function will be executed always on the workers in the same zone of the database.
Lastly, \tapp lets users express a selection strategy for policy blocks. This is
represented by the optional \newFrag{\textit{strategy}} fragment of the
\textit{tag} rule. By default, when we omit to define a \textit{strategy} policy
for blocks, \tapp allocates functions following the blocks from top to
bottom---i.e., \hlopt{best\_first} is the default policy. Here, for example,
setting the \hl{strategy} to \hlopt{random} captures the simple load-balancing
strategy of uniformly distributing requests among the available controllers.

% We then 
%   \item Then, the contributions regarding the topology-based Invoker distribution are listed; this required changes to the Load Balancer and also the use of the previously mentioned \textit{watcher} service.

\subsection{Case Study}
\label{sec:running_example}
As a final illustration of the \tapp language, we show and comment on the salient parts of a \tapp script---reported in \cref{fig:app_example}---that captures the scheduling semantics of the case in \cref{fig:example}.

%\todo{GIGIO: in questo paragrafo appare \#edge; non mi ricordo
%se e' stato gia' introdotto. Sono le etichette dei worker negli edge
%devices? Se si' si puo' dire esplicitamente}
In the script, at lines 1--6, we define the tag associated to \code{critical}
(\critFn) functions: only \codeV{LocalCtl\_1} can manage their scheduling, they
can only execute on \textsf{\#edge}/\codeV{edge} workers (\(W_1,\ldots,W_i\) in
\cref{fig:example}), and no other policy can manage them
(\code{followup:}\codeV{fail}). At line 5 we specify to evenly distribute the
load among all \codeV{edge} workers with \code{strategy:}\codeV{random}.

At lines 7--12, we find the tag of the \code{machine_learning} (\cloudFn{})
functions. We define \codeV{CloudCtl} as the controller and consider all
\textsf{\#cloud} workers (\(W_{k+1},\ldots,W_{j}\) in \cref{fig:example}) as
executors.
% \footnote{Given the deployment schema from \cref{fig:example}, having
% \codeV{*} instead of \codeV{*cloud} at line 10 of \cref{fig:app_example}
% preserves the same semantics. Interestingly, \codeV{*cloud} has the benefit of
% preventing unexpected changes in the semantics due to possible misalignments
% with the deployment schema. For example, if a user mistakenly registers some
% \textsf{\#internal} workers to the \textit{CloudCtl} controller, \codeV{*} would
% allow \textsf{\#internal} workers to run machine-learning functions. Contrarily,
% \codeV{*cloud} averts this risk by restricting valid workers to
% \textsf{\#cloud}-tagged ones.}
%, i.e., any worker in the public cloud
%$W_{k+1},\ldots,W_j$. 
Notice that at line 12 we specify to use the
\codeV{default} policy as the \code{followup}, in case of failure. The
interaction between the \code{followup} and the \code{topology_tolerance} (line
11) parameters makes for an interesting case. Since the
\code{topology_tolerance} is (the) \codeV{same} (zone of the controller
\codeV{CloudCtl}), we allow other controllers to manage the scheduling of the
function (in the \code{default} tag) but we continue to restrict the execution
of machine-learning functions only to workers within the \codeV{same} zone of
\codeV{CloudCtl}, which, here, coincide with \textsf{\#cloud}-tagged workers.

Lines 13--24 define the special, \code{default} policy tag, which is the one
used with tag-less functions (here, our generic ones \genFn{}) and with failing
tags targeting it as their \code{followup} (as seen above, line 12).
In particular, the instruction at line 24 indicates that the \code{default}
policy shall \codeV{random}ly distribute the load on both worker
blocks (lines 14--20 and 21--23), respectively controlled by \codeV{LocalCtl\_1}
and \codeV{LocalCtl\_2}.
Since the two blocks at lines 14--20 and 21--23 are the same, besides the \code{controller} parameter, we focus on the first one. There, we indicate
two sets of valid \code{workers}: the \textsf{\#internal} ones (line 16,
\(W_{i+1},\ldots,W_{k}\) in \cref{fig:example}) and the \textsf{\#cloud} ones
(as seen above, for lines 9--10). The instruction at line 20
(\code{strategy:}\codeV{best\_first}) indicates a precedence: first we try to
run functions on the \textsf{\#local} cluster and, in case we fail to find valid
workers, we offload on the \textsf{\#cloud} workers---in
both cases, we distribute the load \codeV{random}ly (lines 17 and 19).

\section{\tapp in OpenWhisk} \label{sec:tapp_openwhisk}

%In this section, we present how w
We modified OpenWhisk to support \tapp-based
scheduling. In particular, to manage the deployment of
components, we pair OpenWhisk with the popular and widely-supported container orchestrator
Kubernetes\footnote{\url{https://kubernetes.io/}}.
% , due to its popularity and wide support in all
% the major cloud providers.

\cref{fig:openwhisk} depicts the architecture of our OpenWhisk extension, where we reuse the \textit{Workers} and the \textit{Kafka} components, we modify
\textit{Nginx} and the \textit{Controllers}
(\colorbox{LightBlue2}{light blue}\vspace{-.2em} in the picture), and we
introduce two new services: the \textit{Watcher} and the \textit{NFS Server}
(in the \newFrag{highlighted} area of \cref{fig:openwhisk}).

The modifications mainly regard letting Nginx and Controllers retrieve and
interpret both \tapp scripts and data on the status of nodes, to forward requests
to the selected controllers and workers. Concerning the new services, the
Watcher monitors the topology of the Kubernetes cluster and collects its current
status into the NFS Server, which provides access to \tapp
scripts and the collected data to the other components.
Below, we detail the two new services, we discuss the changes to the
existing OpenWhisk components, and describe how the proposed
system supports live-reloading of \tapp configurations. 
We conclude with a description of the deployment procedure applied to the modified OpenWhisk platforms. 
% \todo{adesso c'e' anche la sezioncina sul deployment.. forse e' giusto menzionarla}
% available and supported by all the major cloud providers.

% we first present how we modified OpenWhisk's entry point behavior to handle multiple controller replicas with the addition of a \textit{watcher} service in the deployment platform. We then ...

\subsection{OpenWhisk Controller}
\label{sec:tapp_controller}
To let the original OpenWhisk controller execute \tapp scripts, we extended the existing codebase of OpenWhisk. The component is written in Scala %\footnote{https://scala-lang.org/} 
and 
% to implement our OpenWhisk extension, on top of the original OpenWhisk codebase.
% In particular, the original codebase exposes a
it consists of a base \texttt{LoadBalancer} class
which the vanilla OpenWhisk load balancer extends. To let OpenWhisk support \tapp
scheduling policies, we introduced a new class that also extends the base one, called
\texttt{ConfigurableLoadBalancer}. This class implements a parser and an engine that interpret \tapp scripts.

\vspace{3pt}
\subsection{Watcher and NFS Server Services}
% controller name -> node
% worker name -> node
% worker label -> node
% zone -> node
%
To support \tapp-based scheduling, we need to map \tapp-level information, such as zones and controllers/workers labels, to deployment-specific information, e.g., the name Kubernetes uses to identify computation nodes.
%
% Instead of modifying the existing services to allow them to access deployment-specific properties, for security reasons and generality, we 
The new \textit{Watcher} service fits this purpose: it gathers deployment-specific information and maps it to \tapp-level properties.
%
% Kubernetes  is an open-source system for automating deployment and 
% orchestration of containerized applications. 
To realise the Watcher, we rely on the APIs provided by Kubernetes, which we use to deploy our OpenWhisk variant.
% container orchestrator used to deploy the OpenWhisk platform. 
In Kubernetes, applications are collections of services deployed as ``pods'', i.e., a group of one or more containers that must be placed on the same node
% (e.g. Docker, or any OCI compliant container system)
and share network and storage resources.
Kubernetes automates the deployment, management, and scaling of pods on a distributed cluster and one can use its API to monitor and manipulate the state of the cluster.
% and monitor the pods and other objects already deployed.

% As the former option might present security issues by giving cluster-scoped permissions to externally accessible services (such as Nginx), the latter approach was utilized; the additional service was called \texttt{watcher} in the deployment, and 
Our Watcher
%in this setting can be considered as a simple service performing a periodical query to 
polls the Kubernetes API, asking for pod names and the respective \textit{label}s and \textit{zone}s of the nodes (cf. \cref{fig:openwhisk}), and stores the mapping into the \textit{NFS Server}.
% where they are deployed. 
% The resulting output is parsed using \textit{jq}\cite{web:jq} and written on a JSON file that is stored in a shared volume.
% The resulting output is parsed using \textit{jq}\cite{web:jq} and written on a JSON file that is stored in a shared volume.
% between the Nginx OpenWhisk entry point and the controllers.
% In this way, the service with the highest level of permission in inaccessible from outside, and the other services only interact with it using the shared volume, and reading only a selected subset of the information available via the Kubernetes API.

As shown in \cref{fig:openwhisk}, Nginx uses the output of the Watcher to forward function-execution requests to controllers.
% connect the ``upstreams''---i.e. a group of nodes to forward the requests to--- (named like the controller pods) and the nodes which have their own names and can dynamically host different controller replicas depending on the deployment\todo{Save: cosa vuol dire che Ngnix ``connette'' gli upstreams? In generale, la frase non è chiara.}. 
This allows \tapp scripts to define which controller to target without the need to specify a pod identifier, but rather use a label (e.g., \hlopt{CloudCtl} in \cref{fig:app_example}). Besides abstracting deployment details, this feature supports dynamic changes to the deployment topology, e.g., when Kubernetes decides to move a controller pod at runtime on another node.

\vspace{3pt}
\subsection{Nginx, OpenWhisk's Entry Point}

OpenWhisk's Nginx forwards requests to all available controllers, following a hard-coded round-robin policy.
% This includes all the controller replicas, using a round-robin load balancing algorithm between them.
To support \tapp, we intervened on how Nginx processes incoming request of function execution.

To do this, we used \textit{njs}\footnote{\url{https://nginx.org/en/docs/njs/index.html}.}: a subset of the JavaScript language that Nginx provides to extend its functionalities.
%
% To allow Nginx to route the messages to the correct controller we extend its functionalities by using a njs script.
% % , a script written in a subset of the Javascript language. 
% Nginx indeed support \textit{njs}\footnote{\url{https://nginx.org/en/docs/njs/index.html}} to interact with requests and responses alongside the server's normal behaviour, performing actions such as security checks and redirects. 
Namely, we wrote a njs plug-in to analyse all requests passing through Nginx. The plug-in extracts any tag from the request parameters and compares it against the \tapp scripts. If the extracted tag matches a policy-tag, we interpret the associated policy, resolve its constraints, and find the related node label. The last step is translating the label into a pod name, done using the label-pod mapping produced by the Watcher service.%, obtaining the desired upstream's name.

Since Nginx manages all inbound traffic, we strived to keep the footprint of the plug-in small, e.g., we only interpret \tapp scripts and load the mappings when requests carry some tags and we use caching to limit retrieval downtimes from the NFS Server.
% ; the variables in the request body are always available to Nginx and can be easily parsed as JSON and used in the script. 
From the user's point of view, the only visible change regards the tagging of requests. When tags are absent, Nginx follows the \code{default} policy or, when no \tapp script is provided, it falls back to the built-in round-robin.

\vspace{3pt}
\subsection{Topology-based Worker Distribution} \label{contributions:sec:topology_invoker_distribution}
% \subsubsection{Motivation}

% To allow user to prioritize the scheduling to the near workers,
% 
% The main reason for this addition was to implement the natural completion of the previously described Nginx modifications; that is, after allowing users to select a specific Controller to handle their invocations, having all Controllers prioritize scheduling on the nearest Invokers, effectively grouping them and allowing users to target Invoker sets instead of single workers.
% 
% The concept of ``nearest Invokers'' 
To associate labels with pods,
we exploit the
% was defined in practice with the use of 
topology labels provided by Kubernetes. These labels are names assigned to nodes and they are often used to orient pod allocation. Labels offer an intuitive way to describe the structure of the cluster, by annotating their zones and attributes. In \cref{fig:openwhisk} we represent labels as boxes on the side of the controllers and workers.

Since OpenWhisk does not have a notion of topology, all controllers can schedule
all functions on any available worker. Our extension unlocks a new design
space that administrators can use to fine-tune how controllers access workers,
based on their topology.
At deployment, DevOps define the access policy used by all controllers.
Our investigation led us to identify four topological-deployment access policies:
\begin{itemize}

	\item the \emph{default} policy is the original one of OpenWhisk, where
	      controllers have access to a fraction of all workers' resources. This policy
	      has two drawbacks. First, it tends to \emph{overload} workers, since
	      controllers race to access workers without knowing how the other
	      controllers are using them. Second, it gives way to a form of
	      \emph{resource grabbing}, since controllers can access workers outside
	      their zone, effectively taking resources away from ``local'' controllers;
	      % ; the topology-based priority is still present, but no complete memory
	      % availability is granted.

	\item the \emph{min\_memory} policy is a refinement of the \textit{default}
	      policy and it mitigates overload and resource-grabbing by assigning only a
	      minimal fraction of the worker' resources to ``foreign'' controllers. For
	      example, in OpenWhisk the resources regard the available memory for one
	      invocation (in OpenWhisk, 256MB). When workers have no controller in their
	      topological zone, or no topological zone at all, we follow the default policy.
	      Also this policy has a drawback: it can lead to scenarios where smaller zones
	      quickly become saturated and unable to handle requests;

	\item the \emph{isolated} policy lets controllers access only co-located workers. This reduces overloading and resource grabbing but accentuates small-zone saturation effects;

	\item the \emph{shared} policy
	      %   is to maintain the original behaviour, but 
	      allows controllers to access primarily local workers and let them access
	      foreign ones after having exhausted the local ones. This policy mediates
	      between partitioning resources and the efficient usage of the available ones,
	      although it suffers a stronger effect of resource-grabbing from remote
	      controllers.
	      %
	      %   This approach ; this is a less flexible approach compared to the previous
	      %   two, and can exacerbate situations where functions are often scheduled in
	      %   smaller zones; nevertheless, it's the safest options in that it allows all
	      %   Load Balancers to have a realistic representation of the available memory on
	      %   their Invokers. Similarly to the previous option, Invokers with no assigned
	      %   Controller are shared between all Load Balancers.
	      %

\end{itemize}

When scheduling functions, controllers follow the policies declared in the available \tapp scripts and access topological information and \tapp scripts in the same way as described for Nginx.
In case no \tapp script is available, controllers resort to their original, hard-coded logic (explained in \cref{sec:preliminaries}) but still prioritise scheduling functions on co-located workers.
\vspace{3pt}
\subsection{Dynamic update of topologies and \tapp scripts}
Since both the cluster's topology, its attributes, and the related \tapp scripts
can change (e.g., to include a new node or a new policy tag), we
designed our \tapp-based prototype to dynamically support such changes, avoiding stop-and-restart downtimes.

To do this, we chose to store a single global copy of the policies into the NFS Server, while we keep multiple, local copies in Nginx and each controller instance. When we update the reference copy, we notify Nginx and the controllers of the change and let them handle cache invalidation and retrieval.
% Moreover, Nginx also automatically scans the remote copy for updates every ten minutes.

% This approach presents a compromise between the tradeoff of reading the \tapp policies directly from the remote server that has the most updated copy and the minimization of auxiliary communications involving the Controllers.

\subsection{Deploying \tapp-based OpenWhisk}
\label{sec:deploy_openwhisk}
% also integrated in the OpenWhisk codebase.

The standard way to deploy OpenWhisk is by using the Docker images available for
each component of the architecture---this lets developers choose the
configuration that suits their deployment scenario, spanning single-machine
deployments, where all the components run on the same node, and clustered (e.g.,
via Kubernetes) deployments, e.g., assigning a different node to each component.
Since we modified the Controller component of
the architecture (see \cref{sec:tapp_controller}), we built a new, dedicated Docker image and published it on
DockerHub\footnote{\url{https://hub.docker.com/r/mattrent/ow-controller}.}, so that it
is generally available to be used in place of the vanilla controller.

In this work, for both reproducibility and reliability, we automatised
all the levels of the deployment steps: the provisioning of the virtual machines (VMs) and both the deployment of Kubernetes and of (our extended version of) OpenWhisk.

We programmatically provision VMs using the Google Cloud Platform
% In particular, we provision eight VMs automatically
via a Terraform\footnote{\url{https://www.terraform.io/}.} script. Since this script is tied to a specific topology, we provide more information on it when describing our experiments in \cref{sec:testing}.

We wrote Ansible\footnote{\url{https://www.ansible.com/}.} scripts instead to automatically
deploy the Kubernetes cluster. Given the VMs where one wants to deploy
Kubernetes on and their designated roles (workers, etc.), our Ansible scripts
configure each VM by installing the dependencies required for Kubernetes, deploy
the control-plane on the designated master VM with the \texttt{kubeadm} tool,
and make the other VMs join the cluster as worker nodes (again with the
\texttt{kubeadm} tool). 

Once the Kubernetes cluster is up and running, we use the
Helm\footnote{\url{https://helm.sh/}.} package from \texttt{openwhisk-deploy-kube}~\cite{rejected_projects},
that we forked to implement a \tapp-specific package for the installation with our custom controller image.
This automatically deploys every component on a Kubernetes cluster and allows
the user to parameterize the configuration of the deployment; specifically, we configure the deployment to select our \tapp-based controller image.

All Terraform and Ansible scripts are publicly available at \url{https://github.com/giusdp/ow-gcp}.

%% file: img/app_syntax.tex
\begin{figure*}[t]
	\hspace{3em}\begin{minipage}{.6\textwidth}
\(
\hfil \textit{policy\_tag} \in \textit{Identifiers} \ \cup \ \{ \texttt{default} \} \)
\hfil \(\textit{label} \in \textit{Identifiers} \)
\hfil \(n \ \in \ \mathbb{N}\) \hfil
\[
\begin{array}{lrl}
		\textit{app} & \Coloneqq & \many{\texttt{-}\ \textit{tag}}
		\\
		\textit{tag} & \Coloneqq & \textit{policy\_tag} \ \hlstr{:} \ 
			\many{\texttt{-} \ \newFrag{\textit{controller}?} \quad \textit{workers} \quad \textit{strategy}? \quad \textit{invalidate}?} \quad \textit{strategy}?\quad \textit{followup}?
		\\
		\textit{controller} & \Coloneqq & \newFrag{\hl{controller} \hlstr{:} \textit{label} \quad
																						(\; \hl{topology\_tolerance} \ \hlstr{:}\ ( \hlopt{all} \Div \hlopt{same} \Div \hlopt{none} )\; )?}
		\\
		\textit{workers} & \Coloneqq & \hl{workers} \hlstr{:} \ \many{\texttt{-}\ \hl{wrk}\ \hlstr{:}\ \textit{label}\quad \textit{invalidate}?}
		\\ & \Div & \newFrag{\(\hl{workers} \hlstr{:} \ \many{\texttt{-}\ \hl{set}\ \hlstr{:}\ \textit{label} \quad \textit{strategy}? \quad \textit{invalidate}? }\)}
		\\
		\textit{strategy} & \Coloneqq & \hl{strategy} \ \hlstr{:} \ (\
																								\hlopt{random}
																								\Div \hlopt{platform}
																								\Div \hlopt{best\_first}
																						\ )
		\\
		\textit{invalidate} & \Coloneqq & \hl{invalidate} \ \hlstr{:}\ (\
																								\hlopt{capacity\_used} \ \hlopt{n\%}
																								\Div \hlopt{max\_concurrent\_invocations} \ \hlopt{n}
																								\Div \hlopt{overload}
																						\ )
		\\
		\textit{followup} & \Coloneqq & 
																						\hl{followup} \ \hlstr{:} \ (\
																								\hlopt{default}
																								\Div \hlopt{fail}
																						\ )
\end{array}
\]
\end{minipage}
\caption{\label{fig:capp_syntax}The syntax of \tapp.\vspace{-1em}} 
%(the extensions from\app are \newFrag{highlighted}).\vspace{-1em}}
\end{figure*}

%% file: img/tapp_example.tex
% !TeX root = ../main.tex

\begin{figure*}[t]
%  \begin{center}
\begin{minipage}{\columnwidth}
\begin{lstlisting}[language=yaml,numbers=left,showlines,backgroundcolor=\color{Gold1!20},mathescape=true]
- critical:
  - controller: ~LocalCtl_1~
    workers:
      - $\hl{set}$: ~edge~
        strategy: ~random~
  followup: ~fail~ 
- machine_learning:
  - controller: ~CloudCtl~
    workers:
      - $\hl{set}$: ~cloud~
    topology_tolerance: ~same~
  followup: ~default~
 \end{lstlisting}
 \end{minipage}\hspace{1cm}
 \begin{minipage}{\columnwidth}
 \begin{lstlisting}[language=yaml, numbers=left, firstnumber=last,backgroundcolor=\color{Gold1!20},mathescape=true]
- default:
  - controller: ~LocalCtl_1~
    workers:
     - $\hl{set}$: ~internal~
       strategy: ~random~
     - $\hl{set}$: ~cloud~
       strategy: ~random~
    strategy: ~best_first~
  - controller: ~LocalCtl_2~
    workers: # same as above
    strategy: ~best_first~
  strategy: ~random~
 \end{lstlisting}
 \end{minipage}
%  \end{center}
 \caption{A \tapp script that implements the scheduling semantics of the case study in \cref{sec:introduction} (\cref{fig:example}).}
 \label{fig:app_example}
 \end{figure*}

%% file: benchmark.tex
% !TeX root = main.tex
\section{Evaluation}
\label{sec:testing}

% \todo{Secondo me ci sta bene un mini paragrafo che ricorda che noi permettiamo
% di customizzare le politiche di scheduling al prezzo di usare un nostro
% controller, che deve oltretutto essere istruito con uno script esterno (da
% ricaricare periodicamente in caso di aggiornamenti) e dei monitoraggi
% aggiuntivo relativamente allo stato delle risorse di calcolo disponibili. Gli
% overhead test servono per valutare l'impatto di questo costo aggiuntivo. Poi
% direi che data-locality e' uno dei contesti in cui sappiamo che customizzare
% lo scheduling puo' permettere di avere miglioramenti di performance, e quindi
% abbiamo dei test che cercano di quantificare questi miglioramenti}
We now evaluate our contribution both qualitatively and quantitatively. In
\cref{sec:qual_eval}, we present an edge-cloud-continuum case study, taken from
the literature, to both demonstrate how one can use \tapp to meet topology-aware
functional requirements and how existing serverless solutions---where no
topological information is used by the function scheduler, like vanilla
OpenWhisk---fail in complying with those requirements. In \cref{sec:eval_quant},
we run quantitative benchmarks, comparing the performance of vanilla OpenWhisk
and our \tapp-based variant. We show that the overhead of running \tapp scheduling
policies is negligible and that, in locality-bound scenarios, custom scheduling
policies reduce function run times.

\begin{figure*}[t]
      \begin{center}
            \includegraphics[width=.8\textwidth]{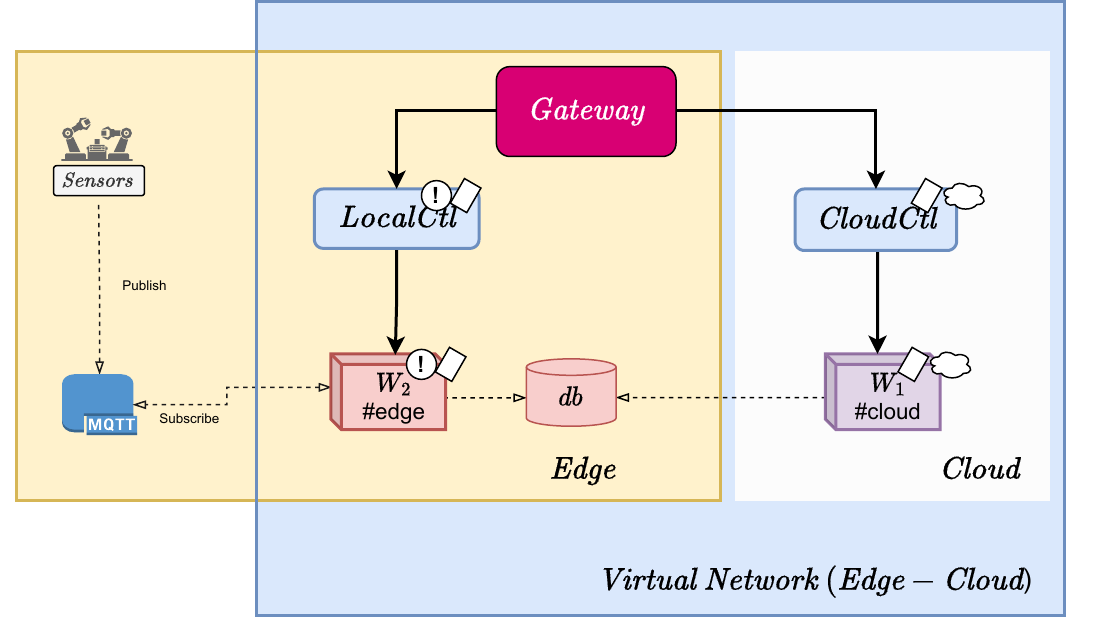}
      \end{center}
      \caption{Representation of the components used in the qualitative evaluation.
            The services were separated in two zones, and connected using two different networks.}
      \label{fig:mqtt_architecture}
\end{figure*}

\subsection{Qualitative Evaluation}
\label{sec:qual_eval}

% First, we set up a qualitative comparison between vanilla OpenWhisk and our
% prototype, to illustrate a realistic case study where the former would lack
% the necessary capabilities to perform the task at hand. In particular, this
% scenario illustrates the importance of topology awareness when scheduling
% serverless functions.

The case study we consider is a simplification of the architecture
described in \cref{sec:introduction}, which we depict in
\cref{fig:mqtt_architecture}. It is a serverless edge-cloud
deployment
of the system described in~\cite{Hong-etal:RealTimeFaultDiagnosisForPowerTransformers,Hong-Jin-Hai:TransformerWindingFaultDiagnosis},
consisting of a power transformers' anomaly detection
application.
Each power transformer to be monitored is equipped
with six accelerometers that produce data at a frequency of 10kHz.
At every minute, the data produced by the six sensors are collected for one second.
For each set of data produced by one sensor, two features are extracted:
FCA (based on Frequency Complexity Analysis) and DET (based on Vibration Stationarity Analysis).
These features are then combined in two vectors (one with the six FCA features and one with the
six DET features) and classified following machine learning techniques.

In the case study, we assume that the
sensors communicate their data through an IoT-specific protocol to a message
broker accessible only from machines within the same local network, named \emph{Edge}. This prevents
public access to the broker to protect it, e.g., from denial of service attacks.
While data gathering happens locally, the elaboration of the data requires powerful resources. For this reason, when local resources are available, we run these
analyses locally, otherwise, we run them in the Cloud.
% \todo{Bisogna aggiornare l'immagine. Togliere ``Virtual Private Network'',
%       mettere ``Local Network (Edge-Cloud)'' nell'area blu. Se volete, cambiare l'area
%       blu con quella gialla, come quella dello use case. Bisognerebbe seguire la
%       stessa struttura dello use case, portando il Cloud a dx e la local a sx. Inoltre
%       è buono includere nell'immagine il Gateway e controllare che il numero e il nome
%       dei controllers sia quello usato nello script \tapp dello use case. Una volta
%       aggiornata l'immagine, aggiungere riferimenti all'immagine nella descrizione del
%       caso d'uso.}

%This use case is similar to the one presented in Figure \ref{img:use-case_excerpt}, with lower resource requirements and complexity, as the former depended on the
%computational resources of the industrial partner, which were not available for personal use in these experiments.
%\todo{Matteo: da rivedere questa spiegazione dello use case}

% In the system described in~\cite{Hong-etal:RealTimeFaultDiagnosisForPowerTransformers,Hong-Jin-Hai:TransformerWindingFaultDiagnosis},
%(i.e. ten thousand of data items produced per second, per sensor).

In our serverless deployment of the use case, we have assumed that the sensors
communicate their data via a standard IoT protocol, namely MQTT~\cite{web:mqtt},
and that the workflow is implemented as a pipeline of three separate functions:
\begin{itemize}
      \item \textbf{data-collection}, which contacts an MQTT broker and subscribes to six topics (one for each sensor), receives the corresponding data and stores it in a local database;
      \item \textbf{feature-extraction}, which queries the database for the collected data and extracts relevant features;
%       (given that the evaluation of a specific machine learning model was beyond the scope of this experiment, in our implementation no significant elaboration is done and we simply implement a function that
%       waits for a fixed amount of time before returning);
      % about it (in our implementation no significant elaboration is done);
      \item \textbf{feature-analysis}, which receives the extracted features and performs the classification task.
%       (also in this case, our implementation of this function simply waits a fixed amount of time before returning).
% %      ,
%            as the evaluation of a specific machine learning model was beyond the scope of this experiment).
\end{itemize}

\begin{figure*}[t]
      \begin{minipage}{\columnwidth}
            \begin{lstlisting}[language=yaml, backgroundcolor=\color{Gold1!20},mathescape=true]
- default: 
  - $\hl{workers}$:
      - $\hl{set}$:	
- MQTT:
  - controller: LocalCtl
    $\hl{workers}$: 
      - $\hl{set}$:
    $\hl{topology\_tolerance}$: $\hlopt{none}$
    $\hl{followup}$: $\hlopt{fail}$
            \end{lstlisting}
      \end{minipage}\hspace{1cm}
      \begin{minipage}{\columnwidth}
            \begin{lstlisting}[language=yaml, backgroundcolor=\color{Gold1!20},mathescape=true]
- DB:
  - $\hl{workers}$: 
    - $\hl{wrk}$: W_2
      $\hl{invalidate}$: $\hlopt{capacity\_used}$ $\hlopt{50\%}$
    - $\hl{wrk}$: W_1 
    $\hl{strategy}$: $\hlopt{best-first}$
- Cloud:
  - controller: CloudCtl
    $\hl{workers}$: 
      - $\hl{set}$: 
    $\hl{topology\_tolerance}$: $\hlopt{none}$
    $\hl{followup}$: $\hlopt{fail}$
            \end{lstlisting}
      \end{minipage}
      \caption{\label{fig:mqtt_tappscript}Configuration used in the MQTT-based experiments.}
\end{figure*}

For this evaluation the platform and services were deployed in two different zones, as represented in Figure \ref{fig:mqtt_architecture}:
a \textit{cloud} zone containing the Kubernetes master node, one OpenWhisk controller and one worker,
and an \textit{edge} zone containing the MQTT broker, the database, one OpenWhisk controller and another worker.
Moreover, the MQTT broker was only accessible from the local network, while the database was accessible from the entire cluster,
and accordingly, from both workers.

In total 6 separate virtual machines were used. For the \textit{cloud} zone, 3 e2-medium instances (2 vCPUs, 4GB of RAM)
were employed on Google Cloud Platform, located in the Belgian data-center (europe-west1-b). One of these acted as the Kubernetes master node and the other 2
as the cloud Controller and the cloud worker, respectively. For the \textit{edge} zone, one physical machine was used for the other 3 virtual machines, of which one hosted the
edge Controller, one the edge worker, and one both the database and the MQTT broker. They were all given 4 cores and 4GB of RAM each.
The various virtual machines were connected using two separate networks:
\begin{itemize}
      \item a virtual private network, containing the entire Kubernetes cluster (i.e. Kubernetes master node,
            cloud and edge Controllers, cloud and edge workers) and the database;
      \item a local network, containing the entire edge zone (i.e. edge worker, edge Controller, MQTT broker and database).
\end{itemize}

This reflected the connectivity constraints of the case study. The MQTT
broker was not part of the virtual private network, and it was not reachable
from the cloud worker. All the components could reach the database. The broker
and the database were not part of the Kubernetes cluster, which only hosted the
OpenWhisk deployment.

The sensors were simulated by a Python script running on the latter virtual machine. To reflect the real system that inspired our experiments,
the workflow frequency was kept at one invocation per minute, and the push frequency of the sensors
was kept at 10kHz (i.e. ten thousand tuples pushed per second, per sensor).
Given that the evaluation of the specific machine learning model used for the analysis was beyond the scope of this experiment, in our implementation no significant elaboration is done and we mocked-up the functions for future extractions and analysis.

In our tests, vanilla OpenWhisk failed every invocation because it tried to
schedule the \textbf{data-collection} function on the cloud worker and failed to
connect to the MQTT broker every time. This comes from the fact that, without
\tapp, OpenWhisk always tries to schedule a specific function on the same worker,
if it is available, and the first worker chosen for the function depends on
the deployment. Since the cloud worker was available and was simply unable to
reach the given service, this was considered a function error by the
scheduler, which caused it to keep scheduling the function on the same node in
the following iterations.

To overcome this limitation, we used \tapp to redirect the different functions to
appropriate zones, as shown in Figure~\ref{fig:mqtt_tappscript}. Specifically:
\begin{itemize}
      \item \textbf{data-collection} (\critFn{}) function was tagged with
      \textit{MQTT} in order to be scheduled only in the \textit{edge} zone, on
      the \textit{LocalCtl} controller, which means on the worker located in the
      same network as the MQTT broker. In case of a scheduling failure, the
      topology tolerance set to \textit{none} forbids the forwarding to the
      other controller, and the invocation stops;
      \item \textbf{feature-extraction} (\genFn{}) function was tagged with
      \textit{DB}. In this case, no controller was defined, but a list of
      workers was given where the worker in the \textit{edge} zone is always
      picked first. Since the database was still accessible from the other
      worker, it is still possible to send the invocation to it in case the
      \textit{edge} worker capacity reaches 50\%. If scheduling fails, it is
      retried one more time with the \textit{default} tag as there is no
      follow-up specified;
      \item \textbf{feature-analysis} (\cloudFn{}) function was tagged with
      \textit{Cloud}. Similarly, for the data-collection function, we gave it a
      specific controller, \textit{CloudCtl}, and a tolerance of \textit{none},
      to use the \textit{cloud} zone exclusively. In this way, we can simulate a
      situation where machine learning tasks would be moved away from the edge
      machines, which usually have fewer resources or stricter requirements.
\end{itemize}

Using this configuration, all invocations performed on our prototype ran
successfully.

\subsection{Quantitative Evaluation}
\label{sec:eval_quant}
We now describe the comparison of the vanilla version of OpenWhisk against our
extension.

As mentioned, \tapp and our \tapp-based extension allows users to overcome the
limitations imposed by state-of-the-art serverless platforms, by defining custom
scheduling policies. In particular, we showed (cf. \cref{sec:tapp_openwhisk})
how we built our platform to support advanced, dynamic modifications of both
policies and of the topology of the machines that run the functions.

Of course, this flexibility comes at the price of adding some complexity to the
serverless platform---e.g., watchers continuously monitor the infrastructure and
update the \tapp scripts at runtime, controllers and gateways execute the custom
logic of the policies at each function invocation, also performing the dynamic
retrieval and application of updated policies.

Hence, to tackle C6 (cf., \cref{sec:introduction}),
% the benchmarks we selected for the comparison and the relevant implementation
% details for running the tests and collecting the logs.
we devise two kinds of tests.
The first kind of tests measure the overhead of the advanced features introduced
by our prototype against the performance of vanilla OpenWhisk. The purpose of
these ``overhead'' tests is to empirically quantify the impact of performance of
the advanced, dynamic features introduced by \tapp in our prototype w.r.t.
vanilla OpenWhisk. Since we want to focus on the performance of the platform,
rather than the execution of the functions, we avoid tests that can introduce
biases generated by data locality effects, i.e., those coming from the
possibility of vanilla OpenWhisk to accidentally choose workers with a
high-latency access to some data source.
The second kind of tests focuses on ``data-locality'' effects and benchmarks the
performance gain of topology-aware policies. The idea with these tests is to
evaluate the performance gains that \tapp-based policies can provide, compared
against the possible suboptimal scheduling of the vanilla version.

In the following description, we
mark (\textbf{O}) overhead tests and (\textbf{D}) data-locality ones.

To perform a comprehensive comparison, we collected a set of representative
serverless test applications, divided into ad-hoc and real-world ones. Ad-hoc
tests stress specific issues of serverless platforms.
% , e.g., cold starts for
% heavy functions, data locality for large queries, and scale-to-zero for idling
% functions {\bf serve dire cosa vuol dire cold-start e scale-to-zero?}. 
Real-world tests are functions taken from publicly-available,
open-source repositories of serverless applications used in production and
selected from the Wonderless~\cite{Eskandani-S:Wonderless} serverless benchmark
dataset.

\paragraph*{Ad-hoc Tests}
%Each 
Ad-hoc tests focus on %a 
specific traits:

\begin{itemize}

      \item \textbf{hellojs} (\textbf{O}) implements a ``Hello World'' application. It provides
            an indication of the performance functions with a simple behaviour which parses and evaluates some parameters and returns a string;

      \item \textbf{sleep} (\textbf{O}) waits 3 seconds. This test benchmarks the handling of
            multiple functions running for several seconds and the management of their queueing process;

      \item \textbf{matrixMult} (\textbf{O}) multiplies two 100x100 matrices and returns the
            result to the caller. This test measures the performance of
            handling functions performing some meaningful computation;

      \item \textbf{cold-start} (\textbf{O}) is a parameter-less variant of \textbf{hellojs} that
            loads a heavy set of dependencies (42.8 MB) required and instantiated when the
            function starts. Moreover, we throttle the invocations of this function to one
            every 11 minutes to let caches timeout\footnote{The default OpenWhisk cache timeout is 10 minutes.}. This function deliberately
            disregards serverless development best practices and showcases how the
            platform handles the cold start of ``heavy'' functions, i.e., the delay
            occurring when a function's code has to be initialised on the worker;

            % \item \textbf{scale-to-zero} is a variant of \textbf{cold-start} that
            % throttles function invocations to one every 11 minutes. In this case, we test
            % how the platform handles infrequent function invocations, assessing the impact of caching, resource wasting, and cache invalidation of function seldom invoked;

      \item \textbf{mongoDB} (\textbf{D}) stresses the effect of data locality by executing a
            query requiring a document from a remote MongoDB database. The requested
            document is lightweight, corresponding to a JSON document of 106 bytes, with
            little impact on computation. This test focuses on the performance of
            accessing delocalized data;

      \item \textbf{data-locality} (\textbf{D}) encompasses both a memory- and bandwidth-heavy
            data-query function. It requests a large document (124.38 MB) from a
            MongoDB database and extracts a property from the returned JSON. This test
            witnesses both the impact of data locality w.r.t latency and bandwidth occupation.

\end{itemize}

All ad-hoc tests use Javascript and run on Node.js 10. The tests that use MongoDB (version 5) run on Node.js 12\footnote{\url{https://github.com/apache/openwhisk-runtime-nodejs}.}.

\paragraph{Real-world Tests}
We draw our real-world tests from Wonderless~\cite{Eskandani-S:Wonderless}: a recent, peer-reviewed dataset that contains almost 2000
projects automatically scraped from GitHub. The projects target %include 
%  by searching for 
% projects developed for
% several serverless platforms~\cite{web:serverless}.
%  The dataset was created to be a source for further research in the serverless
%  ecosystem and it provides examples targeting the 
serverless platforms like AWS, Azure, Cloudflare, Google, Kubeless, and
OpenWhisk.

% The main advantage of using such a dataset is that the code it contains comes from actual repositories; this allowed us to extend the test suite with existing code that has a practical use, instead of developing ad-hoc applications (which may end up being very similar to the basic scenarios).

Wonderless somehow reflects the current situation of serverless industrial
adoption: the distribution of projects in Wonderless is heavily skewed towards
AWS-specific applications. Indeed, out of the 1877 repositories in the dataset,
97.8\% are AWS-specific. Unfortunately, we needed to exclude from our analysis
these projects, since they often rely on using AWS-specific APIs and would
require sensible refactoring to run on other platforms---one such effort could
be subject of future work but escapes the scope of the current one.
%
%
% older, consituting a very significant majority of the available code; despite
% this, because of the scale of this folder, priority was given to the other
% six, as the main purpose of this phase was to extract a small number of
% realistic use cases from the dataset; a complete analysis of the entire
% dataset is beyond the scope of this thesis, and will be reserved for future
% work.
%
This left us with 66 projects which, unfortunately, sometimes carry limited
information on their purpose and usage, they implement ``Hello Word''
applications, and have deployment problems. Thus, to select our real-world
tests, we followed these exclusion criteria:

\begin{itemize}
      \item the project must have a \texttt{README.md} file written in English with at least a simple description of the project's purpose. This filters out repositories that contain no explanation on their inner workings or a description of the project;

            %   \item the project must have more than 10 immediate subfolders (excluding
            %   hidden and common auxiliary folders, i.e. \textit{img}, \textit{env},
            %   \textit{screenshot}, \textit{doc}, and the respective plurals). The purpose of
            %   this rule is to filter out unorganized projects, especially if they are
            %   composed of many subprojects\todo{maybe say that is it possible that this
            %   rule rules out some interesting projects in the limitation section};
            %   The commonly occuring auxiliary subfolders were not counted in this
            %   filtering, since they do not contribute to the project's complexity, and do
            %   not represent independent subprojects. Both rules were applied automatically
            %   before any further inspection.
            %
      \item the project works as-is, i.e., no compilation or execution errors are
            thrown when deployed and the only modifications allowed for its execution
            regard configuration and environment files (i.e., API keys, credentials, and
            certificates).
            % , and the \texttt{serverless.yml} file (if migrated from
            % platforms different from OpenWhisk)\todo{if this is just a config. file, I'd
            % omit this detail on OpenWhisk}.
            %   No significant changes have to be made to the running code, except for
            %   migration-related issues (such as parsing function parameters).
            %
            The reason for this rule concerns both the validity and reliability of the
            dataset, since fixing execution bugs could introduce biases from the
            researchers and skew the representativeness of the sample;
            % and inspecting the projects' code. Indeed, many of the scraped repositories
            % are not maintained anymore, and as such might present a high number of
            % version compatibility issues and bugs;
            %   ; as a complete debugging and renewal of all projects is beyond the scope of
            %   this thesis, only working samples were used.
            %
      \item the project must not use paid services (e.g., storage on AWS S3 or
            deployment dependent on Google Cloud Functions), which guarantees that the
            tests are generally available and easily reproducible;

      \item the project must represent a realistic use case. These exclude ``Hello World'' examples and boilerplate setups. The project must implement at least a function accepting input and producing an output as a result of either an internal transformation (such as code formatting or the calculation of a complex mathematical expression) or the interaction with an external service.
            This rule filters out all projects which do not represent concrete use cases.

            %   \item .
            %   
            %   This rule is complementary to the first one, and the filtering based on it was performed manually; similarly to other rules, its purpose is to avoid the complete analysis of all projects, limiting the selection to repositories where the purpose of the code was made clear; the English language alone was chosen for consistency with the language used in this work.
            %   
\end{itemize}

% \paragraph{Selection}
% Since even the reduced portion of Wonderless considered was still composed by many different projects, a set of rules was defined for this work in order to filter out a vast majority of the repositories; what follows is a list of the rules used, with Rule 0 being the only one applied automatically, and the respective reason for each of them.

% The three following projects passed the previously described selection; they were taken from the analysed folders, and translated for each of the deployed platforms. No modifications were done to the actual running code, save for changes when parsing the functions' parameters (which are handled differently in the various frameworks).

\noindent The filtering led to the selection of these real-world tests:\footnote{For reproducibility, we provide the list detailing the rejection criteria applied to each of the 63 non-AWS projects we discarded at~\cite{repository}.
      %~\cite{rejected_projects}.
}

\begin{itemize}

      \item \textbf{slackpost} (\textbf{O}), from bespinian/k8s-faas-comparison, is a project
            written in Javascript, run on Node.js 12, and available for different platforms.
            % It contains instructions for the deployment and 
            It consists of a function that sends a message through the Slack API. While not
            complex, it is a common example of a serverless application that acts as the
            endpoint for a Slack Bot;

      \item \textbf{pycatj} (\textbf{O}), from hellt/pycatj-web, is a project written in Python,
            run on Python 3.7, and it requires pre-packaged code to work. It consists of a
            formatter that takes an incoming JSON string and returns a plain-text one, where key-value
            pairings are translated in python-compatible dictionary assignments.
            % The repository contains both the web version and the serverless version of the
            % service. We only adapted and deployed the former. 
            As a sporadically-invoked web-based function, it represents an ordinary use case
            for serverless;

      \item \textbf{terrain} (\textbf{D}), from terraindata/terrain, is a project
            written in Javascript and run on Node.js 12. The repository contains a
            serverless application that stress-tests a deployed backend. The backend is a
            traditional, non-serverless application deployed on a separate machine from the
            test cluster, which works as the target for this stress test. This is a common
            example of a serverless use case: monitoring and benchmarking external systems.
            %Having a fixed-location backend makes this test sensible to data locality.

\end{itemize}

\subsection{Test Environment}
\label{sec:test_environment}
% \paragraph{Test execution with JMeter} As all of the deployed actions were
% easily accessible as web endpoints,
To perform load testing and collect benchmarks we used a well-known and stable
tool: Apache JMeter\footnote{\url{https://jmeter.apache.org/}.}.
% Apache JMeter\cite{web:jmeter} was chosen as the load testing and benchmarking
% tool. ; this open source project was originally designed to test web
% applications, and has subsequently expanded to other functions. The way JMeter
% allows sending a certain amount of requests towards a service, simulating the
% presence of multiple users. with the use of threads; JMeter can also be used
% in a distributed fashion, with a main node coordinating the testing and
% various worker nodes actually sending the requests. JMeter tests are defined
% in XML files containing the \textit{test plans}, which are easily editable
% using the GUI.
%
% Separate test plans were defined for each platform\todo{don and each group of
% test cases; also, 
%
We used JMeter without a graphical interface to dedicate all resources to the
tests and maximise the number of concurrent requests. We
consider as metric the
      % the following metrics:
      % 
      % \begin{itemize}
      %   \item
      {latency}, i.e., the time between the delivery of the request and the
reception of the first response.
%   We report the median, average, and maximum;
%   \item error rate of the requests;
%   \item average {connection time}, i.e., the time needed to establish the
%   connection, including the SSL handshake;
%   \item average {throughput}, i.e., the inverse of the average elapsed time,
%   calculated similarly to the latency but considering the last response
%   received.
% \end{itemize}

% We also include a dedicated one for the \textbf{scale-to-zero} test, which needs measurements on the cluster nodes activity (and not only the invocation performance). In this case, we used the Kubernetes default metrics API\footnote{\url{https://kubernetes.io/docs/tasks/debug-application-cluster/resource-metrics-pipeline/}} to sample CPU and memory usage.
% We compared the CPU and memory usage at three different stages: i) \textit{starting}, at the deployment of the platform, with no precedent function invocation; ii) \textit{idle}, after a scale-to-zero event where workers evicted inactive functions; and iii) \textit{processing}, during the invocation of the function.
% \todo{GIGIO: qui si usa ``scale-to-zero event'' ma non si e' mai detto che 
% cosa e'; secondo me bisogna spiegare meglio. Ad esempio, un revisore potrebbe
% anche chiedersi perche' una richiesta ogni 11 minuti (e non un intervallo diverso)}

% \todo{Save: visto che nei results facciamo vedere solo la latency. Togliamo il testo sulle altre metriche? Jac: done}

\paragraph{Configuration}
The basic configuration for
JMeter to run the ad-hoc tests uses
4 parallel threads (users), with a 10-second
ramp-up time, i.e., the time needed to reach the total number of threads, and
200 requests per user.
%four configurations for
%JMeter to run the ad-hoc tests: a basic one and three adjusted to some
%peculiarities of the tests.
% each of these configurations included multiple Thread Groups, executed
% consecutively (one at a time), with differing numbers of repeated requests and
% parallel users.
%
%The basic configuration uses 4 parallel threads (users), with a 10-second
%ramp-up time, i.e., the time needed to reach the total number of threads, and
%200 requests per user.
%
For some ad-hoc tests, we considered more appropriate
a slight modification of the basic configuration.
For the \textbf{sleep} test we use 25 requests per user since we deem it not
necessary to have a larger sample size as the function has a predictable
behaviour.
The \textbf{cold-start} test is long-running, so we use only 1 user
performing 3 requests; this is enough to
witness the effect on cache invalidation and initialisation times.
Since the \textbf{data-locality} test is resource-heavy, we use only 50 repetitions for each of the 4 users; this is enough to
witness data-locality effects.
% This adapts the test to OpenWhisk's default idling tolerance of 10 minutes,
% making sure the platform deallocates the containers even with unmodified
% settings.

We have a different configuration for each Wonderless test:
% \begin{itemize}
% \item 
\textbf{slackpost} has 1 user, 100 repetitions, and a 1-second pause, to account for Slack API's rate limits;
% \item 
\textbf{pycatj} has 4 parallel users, 200 repetitions, and a 10-second ramp-up time, akin to the default for ad-hoc tests;
% \item 
\textbf{terrain} has 1 user, 5 repetitions, and a 20-second pause, since the task is already a stress test and the amount of parallel computation on the node is high.
% \end{itemize}

For each test, we execute 10 runs, removing and re-deploying the whole platform every 2 repetitions to
avoid benchmarking specific configurations, e.g., bad, random configurations
where vanilla OpenWhisk elects as primary a high-latency worker.
The unique exception is \textbf{cold-start}, which is
a long-running test (see the discussion above), for which we considered
3 runs; these are sufficient to evaluate cache invalidation and the
corresponding initialisation times, which are independent of
decisions taken at deployment time.

% We deem this regimen useful to produce stable results, which strengthens the
% validity of the collected data.\todo{qui abbiamo i risultati che possono
% testimoniare per noi. Direi di dire ``As shown by the results presented in the
% next section, this regimen was useful to understand the variability of the
% tests.''} 

% The only exceptions to the rule above is \textbf{cold-start}, which we perform
% once, due to its long runtime and resource requirements.

\paragraph{Cluster}

We deployed both the vanilla and extended versions of OpenWhisk on a cluster
composed of eight virtual machines distributed across two different regions
(corresponding to two zone labels used in the deployment). The vanilla version
is the one from OpenWhisk's official
repository\footnote{\url{https://github.com/apache/openwhisk-deploy-kube}.} at
commit 18960f.

As mentioned in \cref{sec:deploy_openwhisk}, we use a Terraform script to
programmatically provision VMs. Specifically we provision six VMs.
%One of the VMs is for the Kubernetes control-plane node. Three VMs, located in
% the Belgium data centre, host respectively one OpenWhisk controller and two
% OpenWhisk workers. The other four VMs, located in pairs in the Oregon and
% Toronto data centres, host one controller and one worker per pair. 
We used a Kubernetes master node (not used as a computation node by OpenWhisk),
along with one controller and one worker in the first region: \textit{France
      Central}. The other controller and its two associated workers were in the second
region: \textit{East US}. All workers are \textit{Standard\_DS1\_v2} Azure
virtual machines, while the Kubernetes master node and both controllers are
\textit{Standard\_B2s} Azure virtual machines.

% From the point of view of the zones, we have the Kubernetes master node (not
% used as a computation node by OpenWhisk), along with one controller and one
% worker in the first zone: \textit{France Central}. The other controller and
% its two associated workers are in the second zone: \textit{East US}\todo{Non mi quadra questa parte con quanto sopra, e.g., l'Oregon è in West US. In più, per il deployment si diceva che usavamo Google Cloud mentre qui viene detto che viene usata Azure}.
% As such, every machine in the cluster was hosted on Azure.

% as depicted in Figure XXX {\bf che figura? c'e' nella tesi?} \todo{mettere figura che visualizza Table \ref{analysis:table:cluster_config}}.
%Every machine in the cluster was hosted on Azure.
% , since the hybrid nature of the cluster was not necessary for the performance analysis (the geographical and latency differences were enough for data locality tests, and the private/public duality of hybrid clusters was not part of the analysed characteristics). The cluster structure is shown in Table \ref{analysis:table:cluster_config}.

We also deployed two machines in the same AWS region (\textit{us-east}): a \textit{t2.micro} EC2 instance for MongoDB and a \textit{t2.medium} EC2 instance for the \textbf{terrain} backend. All machines (both on Azure and AWS) ran Ubuntu 20.04.

To identify the best target for the data-locality tests, we measured the latency between the five (excluding the Kubernetes master node) cluster nodes and the two EC2 instances, which averages at 2ms for machines located in \textit{East US}, and 80ms for machines located in \textit{France Central}. This identified the \textit{East US} nodes as the optimal targets.

The code used to deploy and run the tests is available at~\cite{rejected_projects}.

\begin{figure*}[h]
\begin{center}
\includegraphics[width=.7\textwidth]{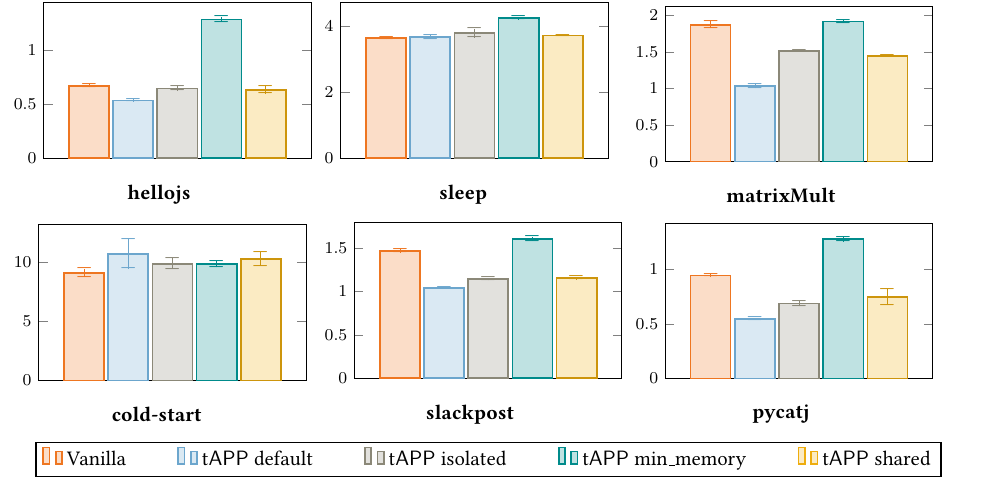}
\end{center}
\caption{Overhead tests (no data locality effects), average latency (bars) and standard deviation (barred lines) in seconds.}
\label{fig:overhead_tests}
\end{figure*}

\begin{figure*}[h]
\begin{center}
\includegraphics[width=.8\textwidth]{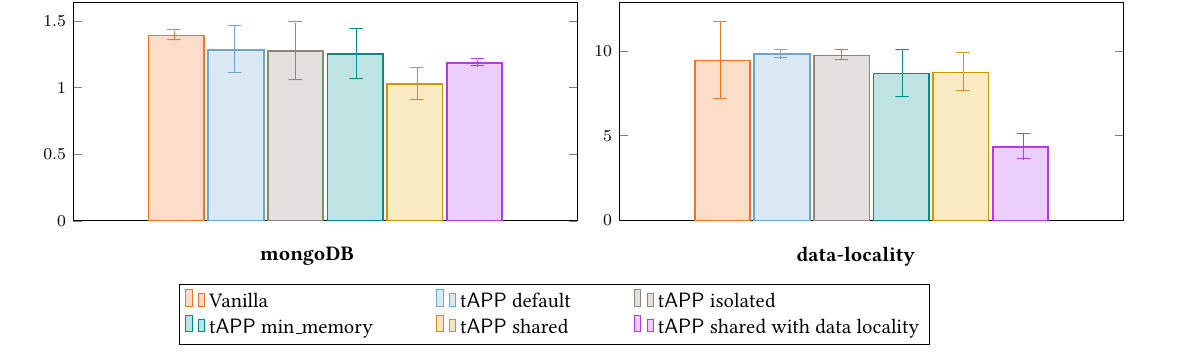}
\end{center}
\caption{Data-locality tests, average latency (bars) and standard deviation (barred lines) in seconds.}
\label{fig:datalocality_tests}
\end{figure*}

\subsection{Results}
\label{sec:results}

We now present the results obtained running our tests on vanilla OpenWhisk
and our topology-aware extension. In particular, we test our extension
under all four topology-based worker distribution policies: \textit{default},
\textit{isolated}, \textit{min\_memory}, and \textit{shared} (cf.
\cref{contributions:sec:topology_invoker_distribution}).

An initial comment regards \textbf{terrain}. While we could deploy this project,
at runtime
%its functions generated 
we observed up to 60\% of timeouts and request errors (in comparison, the other
tests report 0\% failure rate). This test is a real-world one and, according to
our testing methodology (cf. \cref{sec:testing}), we use its code as-is.
% did not edit its code to fix these issues. 
Since this error rate is too high to consider the test valid, we discard it in this section (its raw data is in~\cite{repository}, for completeness).
% ---although, for completeness, we report in \cref{fig:overhead_tests} the
% results from the successful invocations.

% For all the overhead tests
% % , except \textbf{mongo-db} and \textbf{data-locality}, given the
% (free from 
% % data-locality issues 
% data-locality issues), we use no \tapp script.
In the following, we first present the results of the overhead tests and then the results from the data-locality ones.
% (viz., \textbf{mongo-db} and \textbf{data-locality})

% without
% data-locality issues, and tests sensible to data-locality\todo{save: li
% chiamiamo ``overhead'' e ``data-locality'' tests in \cref{sec:testing}.
% Controlliamo di usare gli stessi nomi}. 

\subsubsection{Overhead tests}

% ; in fact, there is no need to schedule the functions on specific workers. 
To better compare the overhead of our extension w.r.t. the vanilla OpenWhisk
one, we run the \textbf{hellojs}, \textbf{sleep}, \textbf{matrixMult},
\textbf{cold-start}, \textbf{slackpost}, and \textbf{pycatj} without a \tapp
script. As a consequence, we also do not tag test functions, since there would
be no policies to run against. As specified in \cref{sec:tapp_openwhisk}, this
makes our platform resort to the original scheduling logic of OpenWhisk,
although it prioritises (and undergoes the overhead of) scheduling functions on
co-located workers. Thus, these tests are useful to evaluate the impact on
performance of our four zone-based worker distribution policies, in comparison
with the topology-agnostic policy hard-coded in OpenWhisk (cf.
\cref{sec:preliminaries}).

We report in \cref{fig:overhead_tests}, in seconds, the average (bars) and the
variance (barred lines) of the latency of the performed tests. For reference, we
report in~\cite{repository} all the experimental data. Since the standard deviation in the results is generally small, we concentrate on commenting the results of the averages.

% given that due to the 
% % extremely 
% high error percentage this test's results fail to offer a significant
% representation of the platform's performance. 

In the results, Vanilla OpenWhisk has better performance w.r.t. all our variants
in the \textbf{sleep} and the \textbf{cold-start} cases, where all tested
policies have similar performance.
%The results are similar for all platforms\todo{deployments? configurations?}
%because these tests have been designed to experiment cases in which 
The latency in these tests does not depend on the adopted scheduling policies,
but on other factors: the three-second sleep in \textbf{sleep}, the long load
times
%  needed to prepare the execution environment 
in \textbf{cold-start}. While we expected a sensible overhead in both cases, we found
% the
% \textbf{cold-start} case the results of \textbf{sleep} are surprisingly
% somehow 
encouraging results: the overhead of
% on queuing processes of our 
topology-based
% customized 
worker selection strategies is
% extremely limited, 
negligible---particularly in the \textbf{sleep}, where the \textit{shared} policy almost matches the performance of vanilla OpenWhisk.

In the other four tests (\textbf{hellojs}, \textbf{matrixMult},
\textbf{slackpost}, and \textbf{pycatj}), the \textit{default} worker
distribution policy outperforms both vanilla OpenWhisk and the other policies.
This policy combines the standard way in which OpenWhisk allocates resources
(where each worker reserves the same amount of resources to each controller) and
our topology-based scheduling approach (where each controller selects workers in
the same zone and uses remote workers only when the local ones are overloaded).
These results confirm that the latency reduction from topology-based scheduling
compensates (and even overcomes) its overhead---in some cases, the performance
gain is significant, e.g., \textbf{matrixMult} shows a latency drop of 44\%.

We deem the good performance of our extension in these tests (spanning
% ad-hoc tests that include 
simple and more meaningful computation and real-word
applications) an extremely positive result. Indeed, we expected topology-based scheduling to mainly allay
% the 
data locality issues%
% of OpenWhisk \cite{Hendrickson-etal:ServerlessComputationOpenLambda}
, but
we have experimentally observed significant performance
improvements also in tests free from this effect.

% As far as the other strategies are concerned,
% A final remark regards 
% the \textit{min-memory} strategy,
% that was conceived to reduce to the minimum the possible resource 
% under utilization of the \textit{default} strategy, which
% statically pre-allocates worker resources to all controllers,
% also to those in different zones.
We also note that the \textit{min\_memory} policy
tends to perform the worst.
To explain this fact, we draw attention to also the results of the \textit{isolated} policy:
% The most likely explanation 
% for this particular result can be obtained with a comparison with the \textit{isolated} policy. 
both strategies can lead to saturated zones when faced with many requests, but they act differently with overloaded local workers.
%in the targeted Controllers' zones are unable to execute the required function. 
%
The \textit{isolated} policy ignores remote workers and returns control to Nginx, which passes the invocation to a different controller. The \textit{min\_memory} policy instead tries to access remote workers with minimal resource availability, which can lead to higher latencies due to queuing and remote communications.
%waiting for the workers' response.
The results of \textit{default} and \textit{shared} reinforce this conclusion: they increase resource sharing within the cluster and mitigate possible asymmetries (here, we had two workers in one zone and one in the other).% which can overload the smaller zone.

\subsubsection{Data-locality tests}

% Before starting to discuss the effects of data locality, as done for the
% results for the overhead tests,
For the data-locality tests, we first comment on the tests we ran following the
same modality of the overhead tests: we do not tag functions and provide no \tapp
script. This lets us compare vanilla OpenWhisk and our extension on a common
ground, where the main difference between the two stands on the four
distribution policies applied at deployment level and their overhead. Then, we
ran the same tests (on our extension), but we tagged the functions and provided
a \tapp script that favours executing functions on workers close to the
data source.

We report in \cref{fig:datalocality_tests}, in seconds, the average (bars) and
the variance (barred lines) of the latency of the data-locality tests
\textbf{mongoDB} and \textbf{data-locality}---the full experimental data is in~\cite{repository}. For brevity, we show, with the right-most bar in
\cref{fig:datalocality_tests}, the results of the best-performing distribution
policy (\textit{shared}, see below) paired with the mentioned \tapp script.

As expected, in all tests our extension outperforms vanilla OpenWhisk,
confirming previous evidence on data
locality~\cite{HSHVAA16}
and presenting useful applications of topology-aware scheduling policies for
topology-dependent workflows.

In \textbf{mongoDB}, our extension outperforms vanilla OpenWhisk under all
strategies, although it undergoes a higher variance. The small variance of
vanilla OpenWhisk in this test is probably thanks to the light test query, which
mitigates instances where vanilla OpenWhisk uses high-latency workers.

The results from \textbf{data-locality} confirm the observation above. There,
the variance for vanilla OpenWhisk is larger---quantitatively, the variances of
\textbf{mongoDB} for our extension stay below 0.5 seconds, while the variance of
vanilla OpenWhisk in \textbf{data-locality} is 9-fold higher: 4.5 seconds. Here,
the heavier test query strongly impacts the performance of those ``bad''
deployments that prioritise high-latency workers.

More precisely, the best performing strategies are \textit{shared} for
\textbf{mongoDB} and \textit{min\_memory} for \textbf{data-locality}. In the
first case, since the query did not weigh too much on latency (e.g.,
bandwidth-wise), mixing local and remote workers favoured the \textit{shared}
policy, which, after exhausting its local resources, can freely access remote
ones. In the second case, the \textit{min\_memory} policy performed slightly
better than the \textit{shared} one. We attribute this effect to constraining
the selection of workers mainly to the local zone and resorting in minimal part
to remote, higher-latency workers.

% thus maximizing the
% number of function execution in its zone. 
% This interpretation of the results is confirmed
% by the fact that also the \textit{min-memory} strategy
% --the worst one in most of the overhead tests ---has good performances in the data locality tests.
% Indeed, \textit{min-memory} characteristic is to allow controllers to use almost 
% all of the local resources, given that only a minimal amount 
% of them is dedicated to remote controllers.

Given the results above, we performed the \tapp-based tests (right-most column of \cref{fig:datalocality_tests}) with the
\textit{shared} policy\footnote{In \textbf{data-locality},
      \textit{min\_memory} has a slightly lower average than \textit{shared}, but the latter has both lower variance and maximal latency.}.
% Since these function should access to external data, we used a \tapp script to
% preferably select workers located on machines in the Azure region \textit{East
% US}, having lower-latency access to the database deployed in the AWS
% \textit{us-east} region.

% For these benchmarks sensible to data locality,
% we have decided to include an additional experiment in which
% we 
% avoid enriching the data and therefore not use any information on data locality.
% % exploit locality also at the level of controllers.
% This was obtained by repeating the two tests
% adopting the best performant \textit{shared} strategy,
% with the indication to Nginx to use the controller 
% in \textit{East US} as the primary controller,
% and the one in \textit{France Central} as a secondary one.
% This was possible simply by enriching 
% the \tapp script with the explicit indication of the main
% controller and the adoption of the 
% topology tolerance \texttt{all}, allowing the infrastructure
% to move to other controllers when the main one is not responsive.

% The results of the tests using the share strategy final tests are reported in
% the last colum in the plots of \cref{fig:datalocality_tests}.
% ,
% on the right of the two bars depicting the performances of the \textit{shared}
% strategy. 
Compared to the tag-less \textit{shared} policy, the tagged case in \textbf{mongoDB} is a bit slower, but more stable (small variance). In
\textbf{data-locality} it almost halves the run time of the tag-less case.

% A final remark is concerned with these final results about
% controller-based locality in the \textbf{mongoDB} case.
% Including controller-based locality requires Nginx to 
% access external resources (i.e., the \tapp script) to 
% decide the controller to use. This could be a significant
% overhead for a component that must be highly performant.

%Interestingly, t
These tests witness the trade-off of using \tapp-based
scheduling to exploit data locality and the overhead
of parsing the \tapp script: due to its many lightweight
requests, \textbf{mongoDB} represents the worst case for
the overhead, but the test still outperforms vanilla OpenWhisk
(showing that the overhead is %more than 
compensated by the advantages of our worker selection
strategies);
%performs many requests for a lightweight payload, which does not compensate for
%the overhead of parsing and computing scheduling policies (future work
%can tackle this problem with solvers and other high-performance engines); 
in \textbf{data-locality}, the heaviness of the query and the payload
favours spending a small fraction of time to route functions to the workers
with lower latency to the data source.

%% file: related.tex
\section{Related Work}
\label{sec:related}

% We compare with related work on optimisation of serverless function scheduling.
%
% Despite the relatively young age of serverless computing (AWS Lambda was only introduced at the end of 2014\cite{web:IntroducingAwsLambda})
% Lambda was not the first

Many works on serverless focus on minimising the latency of function invocations.
% Thise solution is then evaluated by modifying the Apache OpenWhisk platform,
% and comparing it with an unmodified version, showing significant improvements
% in startup times.
%
%%%%
% Formal models
% Another line of research covers formal methods for serverless, Gabbrielli et al \cite{gabbrielli-etal:NoMoreNoLessFormalModelServerless}
% present the Serverless Kernel Calculus (SKC), a core calculus
% for Serverless which combines ideas from $\lambda$-calculus (for functions) 
% and $\pi$-calculus (for communication); SKC defines a serverless architecture as a pair $\langle S,D \rangle$ where $S$ is the system of running
% functions and $D$ is a definition repository. The authors then use an extension of SKC to encode a portion
% of Tailor\cite{web:TailorAWSAccountProvisioning}, an architecture for user registration developed over AWS Lambda.
% %
% In the same year, a different work \cite{Jangda-P-B-G:FormalFoundationsServerlessComputing} proposed formal tools
% for serverless computing, defining an operational semantics for of serverless platforms called $\lambda_{\bblambda}$;
% this model also captures low-level details such as cold starts, storage, transactions and function restarts.
% The authors then extend $\lambda_{\bblambda}$ with a domain specific language for composing serverless functions,
% called ``serverless programming language'' (SPL), and implement different programs as case studies to identify
% possible features that might be missing from the language.
%
%%%%
% Scheduling
Several of them tackle the problem by optimising function scheduling~\cite{kuntsevich2018distributed,shahrad2019architectural}.

One work close to ours is by Samp\'{e} et
al.~\cite{10.1145/3135974.3135980}, who present an approach that allocates
functions to storage workers, favouring data locality. The main difference with
our work is that the one by Samp\'{e} et al. focusses on topologies induced by
data-locality issues, while we consider topologies to begin with, and we capture
data locality as an application scenario.

More in general, Banaei et
al.~\cite{Banaei-S:ETASPredictiveSchedulingFunctionsWorkerNodes} introduce a
scheduling policy that governs the order of invocation processing, depending on
the availability of the resources they use.
Abad et al.~\cite{abad2018package} present a package-aware technique that
favours re-using the same workers for the same functions to cache dependencies.
Suresh and Gandhi~\cite{SG19} show a scheduling policy oriented by resource
usage of co-located functions on workers.
Steint~\cite{stein2018serverless} and Akkus et al.~\cite{akkus2018sand}
respectively present a scheduler based on game-theoretic allocation and on the
interaction of sandboxing of functions and hierarchical messaging.
Other scheduling policies exploit the state and relation among functions. For
example,
% State management in serveless application is also a
% challenge~\cite{Schleier-Smith-etal:WhatServerlessComputingIsAndShouldBecome,Hassan-etal:SurveyOnServerlessComputing}.
%
% The literature contains works towards  as well, such as
Kotni et al.~\cite{Kotni-N-G-B:FaastlaneAcceleratingFaaSWorkflows} present an
approach that schedules functions within a single workflow as threads within a
single process of a container instance, reducing overhead by sharing state among
them.
% , as functions can simply exchange data via simple load/store instructions;
% this represents both a study on optimizing function scheduling and on handling
% the issue of managing state in a specific workflow. have recognised it both as
% a challenge for the paradigm and a difficult task for developers; the
% aforementioned
Shillaker and Pietzuch~\cite{Shillaker-P:FaasmLightweightIsolation} use state by
supporting both global and local state access, aiming at performance
improvements for data-intensive applications.
Similarly, Jia and
Witchel~\cite{Jia-W:BokiStatefulServerlessComputingSharedLogs} associate each
function invocation with a shared log among serverless functions.
%
% Another example of a stateful FaaS runtime is
% Cloudburst\cite{Sreekanti-etal:CloudburstStatefulFaaS}, a framework built on
% top of the Anna\cite{Wu-etal:AnnaKVSForAnyScale} key-value storage (KVS).
%
% a performance comparison between Boki and Cloudburst is also provided by the
% first runtime's authors in their paper.
%
% IoT and Edge

An alternative approach, besides improving scheduling, is
reducing the number of cold starts when launching new
functions~\cite{Jonas-etal:BerkeleyViewOnServerless,Hellerstein-etal:ServerlessOneStepForwardTwoStepsBack}.
%
% One of the main problems discussed is the cold start, i.e., the delay that
% functions invocation incur to be executed if their code is not yet initialize
% on the workers node.
In this direction, Shahrad et al.~\cite{SFGCBCLTRB20} introduce an
empirically-informed resource management policy that mediates cold starts and
resource allocation. 
% Solaiman et
% al.~\cite{Solaiman-A:WLECNotSoColdArchitectureMitigateColdStartServerless},
% present an architecture that uses queues to ``pre-warm'' and cache containers,
% reducing startup times at the cost of moderately higher memory usage.
%
Silva et al.~\cite{Silva-etal:PrebakingFunctionsWarmServerlessColdStart} propose
% a similar compromise 
a solution based on process snapshots: when the user deploys the
function, they generate/store a snapshot of the process that runs that function
and, when the user invokes the function, they load/run the related snapshot.
 % generally reducing the overhead from the snapshot creation.
%
Similarly, on network elements, Mohan et
al.~\cite{Mohan-etal:AgileColdStartsForScalableServerless} present an approach
based on pre-creating networks and connecting them to (a pool of) containers
``paused'' after the network-creation step. At function invocation, they skip
network-initialisation start-up times by attaching to one of the containers and
completing the initialisation from there.

% Serverless has been also studied for IoT and edge computing.
% % , and what challenges they present, is an interesting field as well. A study
% % combining serverless computing with edge and IoT devices is presented in
% For example, Glikson et al.~\cite{Glikson-etal:DevicelessEdgeComputing}
% introduced the concept of ``deviceless edge computing''; a paradigm that allows
% functions to be executed on devices close to the user effectively integrating
% those devices in the serverless infrastructure.
% %
% This is expanded in~\cite{Gusev:ServerlessDevicelessDewComputing}, where the
% possibility of ``infrastructureless'' computing is explored, evaluating
% challenges and advantages of including embedded systems in the paradigm of
% serveless.
% % , while also using concepts from dew computing (executing functions on a local
% % smart device, and offloading it to a server or a neighboring smart device if
% % specific conditions are met, such as the presence of an internet connection or
% % a low battery level); it has to be noted that the author only provides an
% % analysis of this approach, not an implementation.

The main difference between these works and our proposal is that in the former
topologies (if any) emerge as implicit, runtime artefacts and scheduling do not
directly reason on them. Moreover, being a general approach to scheduling, future work on \tapp can include scheduling policies proposed in these works, e.g., as strategies for worker selection.

% \paragraph{Other, open-source serverless platforms}

% For completeness, we briefly review the other two most popular open-source
% serverless platforms, OpenFaaS and OpenLambda, to confirm that they share the
% same concepts and elements present in Apache OpenWhisk.

% % Fission is written in Go and it heavily depends on Kubernetes. Fission has a
% % programming model similar to that of Apache OpenWhisk---based on functions,
% % environments, and triggers (the latter two not described here for brevity).
% % While the components for function invocation and execution mirror those of
% % Apache OpenWhisk\todo{right?}, the functionalities of the related components are
% % all implemented via Kubernetes CRDs (Custom Resource Definitions).

% OpenFaaS uses Kubernetes as scheduler, runtime, and deployment
% platform~\cite{web:faasd}. Functions run inside Docker containers equipped with
% a ``watchdog'' HTTP server that executes them when invoked by the OpenFaaS API gateway. Kubernetes handles the deployment of watchdogs,
% their replication, and load balancing. The platform supports a scale-to-zero
% policy to remove idle watchdogs from workers.

% OpenLambda~\cite{HSHVAA16} exposes function invocations through a load balancer
% (Nginx), which directly selects a worker. When a worker receives a request, it
% runs the function in a Docker container. The first time a function runs on a
% worker, the worker fetches the function definition from a code store and caches
% it for future invocations.

%% file: conclusion.tex
\section{Conclusion}
\label{sec:conclusion}

We introduced \tapp, a declarative language that provides DevOps with finer control on the scheduling of serverless functions. Being topology-aware, \tapp scripts can restrict the execution of functions within zones and help improve the performance (e.g., exploiting data or code locality properties), security, and resilience of serverless applications. 
% Moreover, it allows DevOps to exploit data and session locality to boost the performance of the application and reduce the latency.

To validate our approach, we presented a prototype \tapp-based serverless platform, developed on top of OpenWhisk,
% (the most popular open-source serverless platform(). 
and we used it to
% We have 
show that topology-aware scheduling is on par or outperforms hard-coded, vanilla OpenWhisk scheduling---e.g., tests that stress data locality gain considerable latency reductions.

Future work includes applying \tapp on different platforms, e.g., OpenLambda,
OpenFAAS, and Fission. We plan to expand our range of tests: both to include other aspects of locality (e.g., sessions) and specific components of the platform (e.g., message queues, controllers) and new benchmarks for
alternative platforms, to elicit the peculiarities of each implementation. 
%
% be worth Besides providing further proof of the realisability of
% topology-aware scheduling, different implementations would let us conduct
% performance tests using larger clusters with applications that can put under
% heavy stress the controllers and the system used to dispatch the messages
% (e.g., Kafka in OpenWhisk). 
Regarding tests, we remark the general need for more platform-agnostic and
realistic suites, to obtain fairer and thorough comparisons.
% We think of this as
% a community effort, involved in creating a standard, curated set of
% benchmarks.
% ---e.g., starting from %a refinement of automatically-scraped projects,
% the work done here for Wonderless~\cite{Eskandani-S:Wonderless}.

We also intend to formalise the semantics of \tapp, e.g., building on existing
``serverless calculi''~\cite{GGLMPZ19,JPBG19}. This is a
stepping stone to mathematically reason on scheduling policies and formally prove they provide desirable guarantees.

Finally, we would like to support DevOps in the optimization of their serverless
applications by studying and experimenting with heuristics and AI-based
mechanisms that profile applications and suggest optimal \tapp policies.
Similarly, extensions of \tapp could benefit from interactions with frameworks able to specify function compositions, e.g., Yussupov et al.~\cite{ysbl} recently introduced a method for modelling and deploying serverless function orchestrations which one could use to extract execution dependencies among functions and inform the synthesis of \tapp policies that optimise the overall execution of compositions.

% \todo{Future work, introdurre un linguaggio per specificare i deployment, che possiamo usare per controllare la validata (e altro?) di \tapp scripts.}
% 
% \todo{Future work, profiler per i benchmarks.}

% \todo{add a remark in the conclusions and limitation on the fact that wonderless is not that usable?}